\documentclass[11pt,a4paper]{article}

\usepackage[left=1.0in, right=1.0in, top=1.0in, bottom=1.0in]{geometry}

\usepackage{amssymb}
\usepackage{amsfonts}
\usepackage{amsmath}
\usepackage{graphicx}
\usepackage[font=small]{caption}
\usepackage{subcaption}
\usepackage{bm}
\usepackage[ruled,vlined]{algorithm2e}
\usepackage{algpseudocode}
\usepackage{varwidth}
\usepackage{placeins}
\usepackage[utf8]{inputenc}
\usepackage{mathtools} 
\usepackage{booktabs}
\usepackage{placeins}
\usepackage{verbatim}
\usepackage{enumitem}
\usepackage{sidecap}

\usepackage[]{color-edits}
\addauthor{yk}{red}
\addauthor{nn}{blue}
\newcommand{\remove}[1]{}  
\usepackage[dvipsnames]{xcolor}

\newcommand{\qed}{\rule{1.5mm}{2mm}\vspace{0.1in}}

\newcommand{\ignore}[1]{}

\algtext*{EndFor}
\algtext*{EndIf}

\title{On the Effectiveness of Tracking and Testing in SEIR Models}
\author{
Yoav Kolumbus\\ Hebrew University of Jerusalem \\ {\small yoav.kolumbus@mail.huji.ac.il} \and 
Noam Nisan\\ Hebrew University of Jerusalem \\ {\small noam@cs.huji.ac.il}
}

\renewcommand\footnotemark{}

\date{ 
\thanks{\hspace{-21pt} This project has received funding from the European Research Council (ERC) under the European Union’s Horizon 2020 research and innovation programme (grant agreement No 740282).}
}

\begin{document}
\maketitle

\begin{abstract}
We study the effectiveness of tracking and testing in mitigating or suppressing epidemic outbreaks, in combination with or as an alternative to quarantines and global lockdowns. We study these intervention methods on a network-based SEIR model, augmented with an additional probability to model symptomatic, asymptomatic and pre-symptomatic cases. Our focus is on the basic trade-offs between economic costs and human lives lost, and how these trade-offs change under different lockdown, quarantine, tracking and testing policies.   

\vspace{5pt}
\noindent
Our main findings are as follows:
\begin{itemize}[leftmargin=*]
	\item Tests combined with patient quarantines reduce both economic costs and mortality, but require a large-scale testing capacity to achieve a significant improvement.  
	\item Tracking significantly reduces both economic costs and mortality.
	\item Tracking combined with a limited number of tests can achieve containment without lockdowns.
	\item If there is a small flow of new incoming infections, dynamic ``On-Off'' lockdowns are more efficient than fixed lockdowns.
\end{itemize}
Our simulation results underline the extreme effectiveness of tracking and testing policies in reducing both economic costs and mortality and their potential to contain epidemic outbreaks without imposing social distancing restrictions. This highlights the difficult social question of trading-off these gains with the privacy loss that tracking necessarily entails.    
\end{abstract}

\section{Introduction}
This paper is being written as the COVID-19 corona virus epidemic is still ravaging its way across the
world \cite{world2020coronavirus,covid2020severe,remuzzi2020covid,WHO}. 
While the basic models for the spread of infectious disease like COVID-19 are well understood,
it seems that in the case of COVID-19 we do not have credible estimates for many of the basic 
parameters governing the behavior in such models.  
Examples of these parameters include the rate of infection
($R_0$), the effect of basic types of social distancing on this rate, the fraction of asymptomatic
people among the infected, and the degree to which these asymptomatic individuals are infectious.  

The present paper studies the effectiveness of mitigation efforts 
on the spread 
of an infection in the classical SEIR model \cite{kermack1927}.  
We leave the determination of the real-life parameters of COVID-19 to epidemiologists (e.g., \cite{lauer2020incubation,liu2020reproductive,zhang2020estimation,anderson2020will,Science-ferretti2020quantifying}), 
and use values that on the one hand are plausible in the COVID-19 context, and on the other hand
demonstrate our basic qualitative findings.
Our interest in this paper is in the basic forms of controlling the spread of the disease: quarantine 
of infected patients, testing of the population, and tracking the contacts of infected patients, possibly
testing or quarantining them as well.  In order to be able to simulate a situation where contacts of
patients are actually tracked, we use a model of an individual-level network of contacts \cite{newman2002spread,Nature2004Networks,liu2017analysis}, 
rather than a mathematical analysis at a population level \cite{hellewell2020feasibility,glover2020health,prem2020effect,browne2015modeling}.

\subsection{Our Model}
Here are the main features of our model; details appear in Section \ref{sec:model}.
\begin{itemize}[leftmargin=*]
\item As in the standard SEIR model, each individual can be in one of four states: susceptible, exposed, infectious,
or removed.  Except for a few individuals that start as infectious, all individuals start as
{\bf susceptible}.  When a susceptible individual has contact with an 
{\bf infectious} patient, he gets infected with some probability, at which point he moves to an {\bf exposed} state -- 
a non-contagious  
incubation period -- after which he becomes 
contagious 
and moves to an 
{\bf infectious} state.  
Finally, the patient recovers (or, with a certain probability, dies), moving to a {\bf removed} state.  
\item A population of $N$ individuals is represented as vertices in a graph, where two individuals are 
linked by an edge if there is a possible contact between them.\footnote{In our simulations
the set of possible contacts of a vertex was chosen uniformly at random, leading, on average, to uniform mixing within the population as in the classical SEIR model. 
In Section \ref{sec:geo-graphs} in the appendix we use a different graph model that takes into account geographic factors. 
The results under both models are qualitatively similar.} 
Every period (``day'')
each infectious individual interacts with a random subset of his possible contacts and
infects, with some probability, each of them.  
\item A basic parameter that determines how difficult it is to quarantine infectious individuals is whether
they are symptomatic or not.  If all infected individuals are symptomatic and can be immediately quarantined once they enter the
infectious state, then the disease does not spread at all. As we are focused on measures of controlling
the spread under the realistic more difficult settings, 
a critical parameter in our model is 
what fraction of the infected population remain asymptomatic, while still being infectious.
\item We use a single ``lockdown level'' parameter to capture the total effect of all social 
distancing measures combined.
This parameter simply counts the percent reduction in probability of meeting each one of the possible
contacts.  Thus, ``$80\%$-lockdown'' means that the average number of contacts of each 
individual was decreased by a factor of 5.
\end{itemize}

\begin{figure}[t!]
\centering
	\begin{subfigure}{.345\linewidth}
		\includegraphics[width=1.05\linewidth]{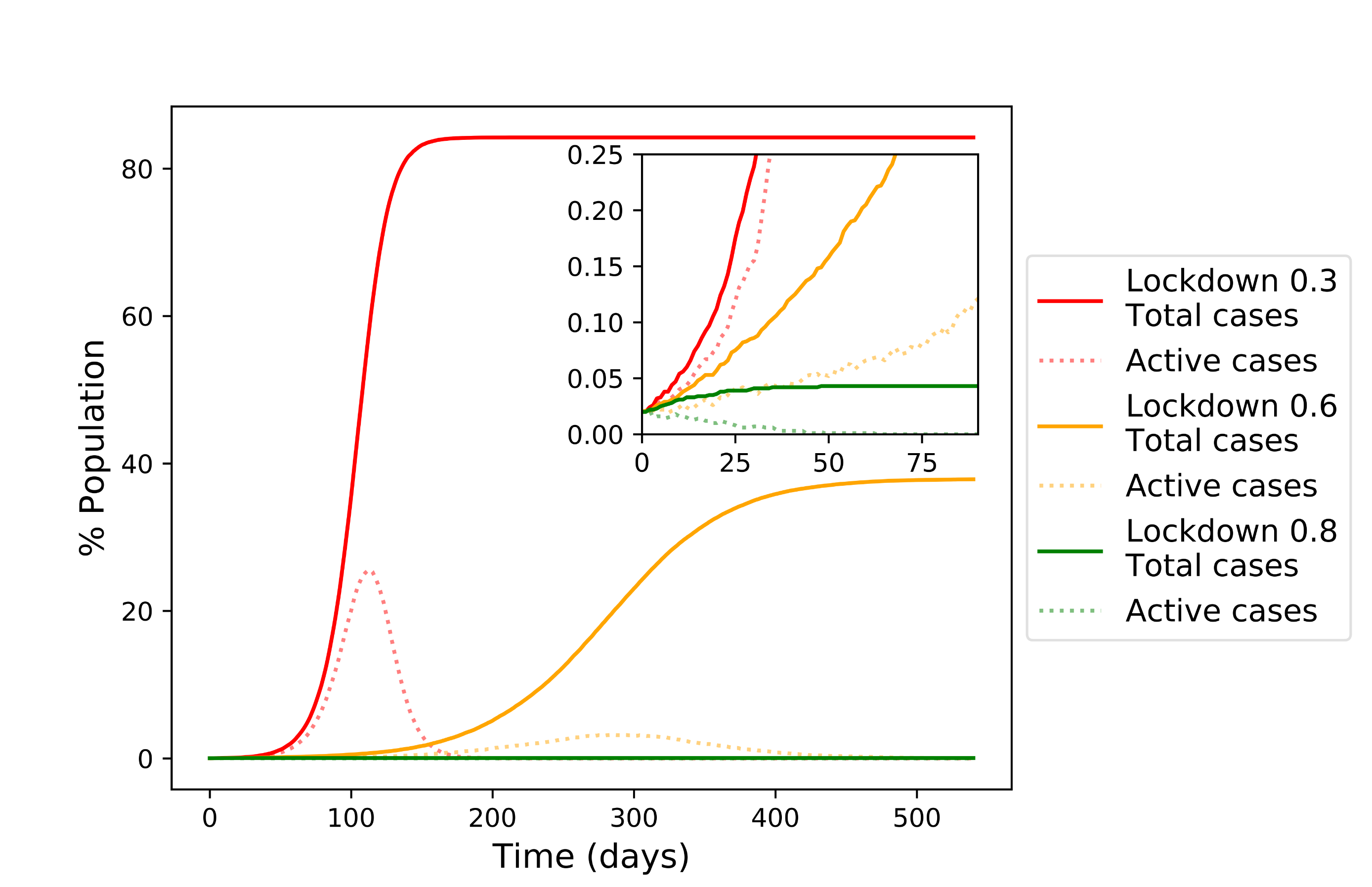}
		\caption{Outbreak dynamics}  
		\label{fig:example_dynamics}
	\end{subfigure}
	\begin{subfigure}{.31\linewidth} 
		\includegraphics[width=1.02\linewidth]{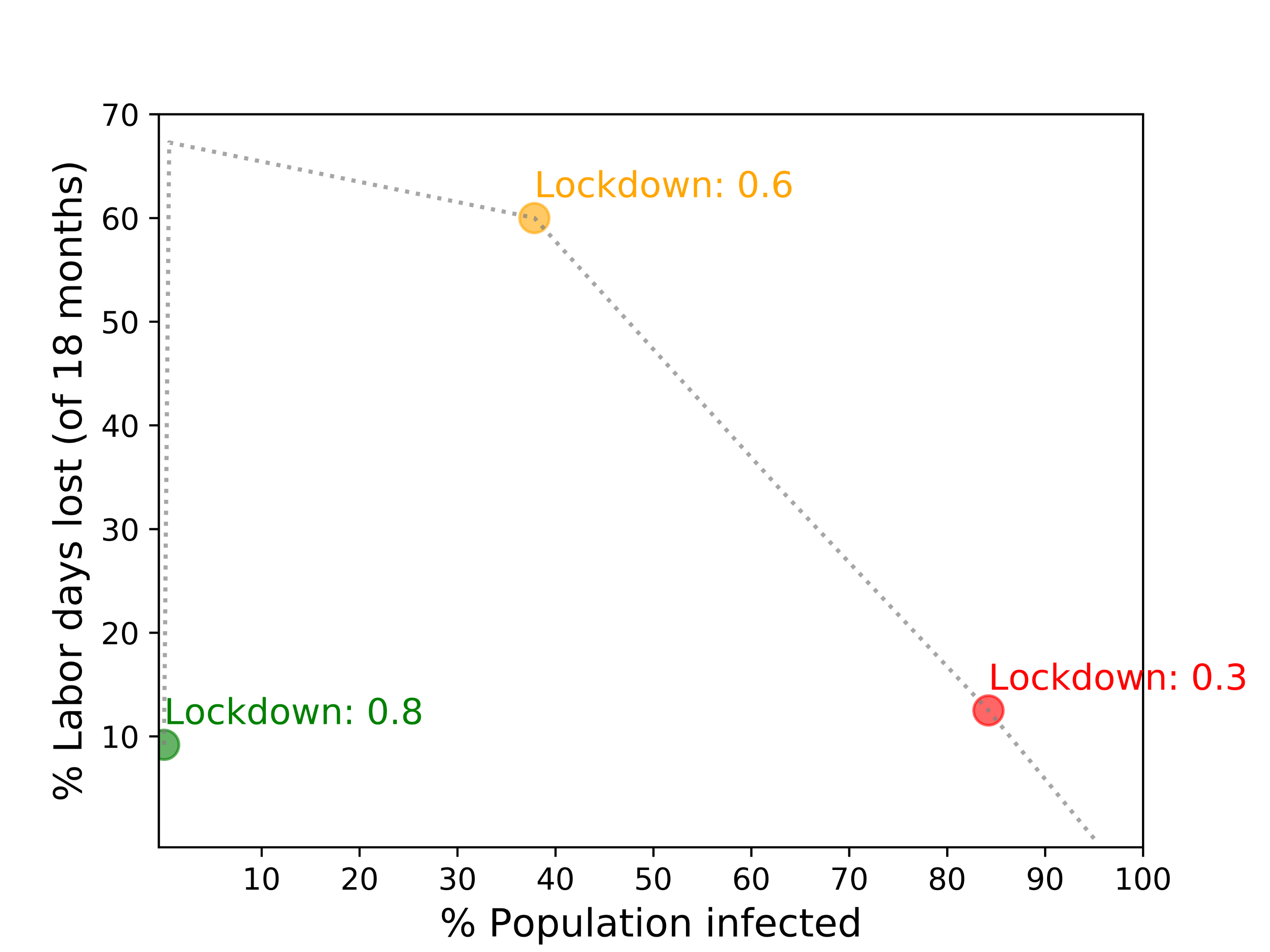}
		\caption{Closed system}  
		\label{fig:example_cost_map}
	\end{subfigure}
		\begin{subfigure}{.31\linewidth} 
		\includegraphics[width=1.02\linewidth]{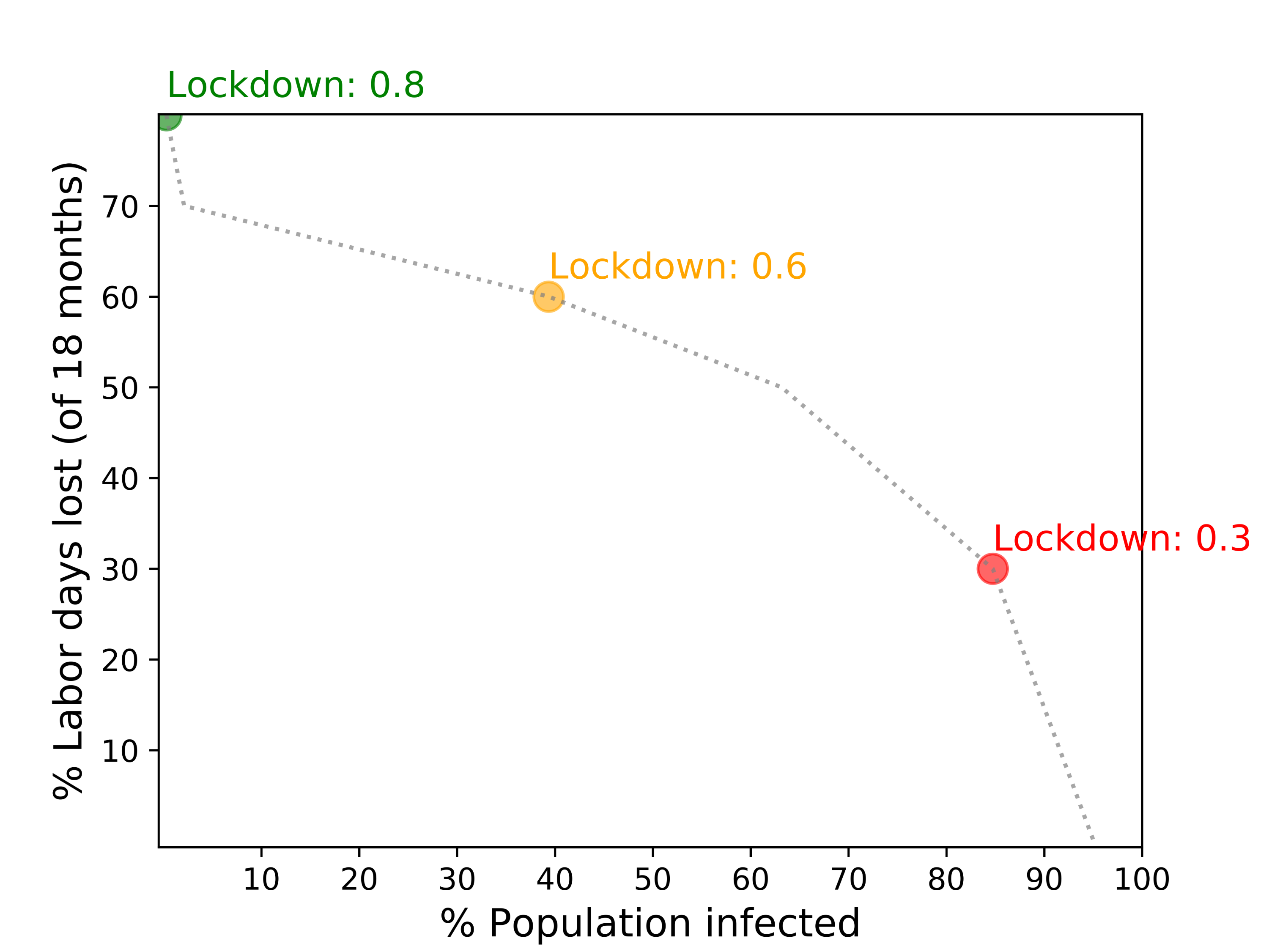}
		\caption{Open system}  
		\label{fig:example_cost_map_with_flow}
	\end{subfigure}
\caption{
Outbreak dynamics and outcomes with different lockdown levels. 
\ref{fig:example_dynamics}: Outbreak dynamics. The full and dotted lines show the total percentage of population infected and the percentage of active cases, respectively, as a function of time. \ref{fig:example_cost_map},\ref{fig:example_cost_map_with_flow}: The cost map of the  percentage of labor days lost in a period of 18 months versus the percentage of population infected, in closed and open systems, respectively. 
}
	\label{fig:example_dynamics_and_cost}
\end{figure}

\noindent
We study the following interventions for controlling the spread of the disease: 
(1) Basic social distancing; 
(2) Quarantine of symptomatic infectious individuals; 
(3) Testing of selected individuals, tests that can discover also exposed and asymptomatic individuals; 
(4) Tracking of an individual's previous contacts, allowing e.g. testing or quarantining them. 
\remove{
\begin{itemize}
\item Basic social distancing.
\item Quarantine for symptomatic infectious individuals.
\item Testing of selected individuals, tests that can discover also exposed and asymptomatic individuals.
\item Tracking of an individual's previous contacts, allowing, e.g. testing or quarantining them. 
\end{itemize}
}

Our main focus is on the trade-offs between the extent of the spread of the disease and 
the economic harm caused by the various restrictions.  For simplicity we quantify the 
former by ``human lives lost'' (a fraction
of the total number of infected) and the later by ``fraction of labor days lost'' due to social distancing or
quarantine.\footnote{For simplicity we assume that the economic loss in a day with a lockdown level of 
$X\%$ is proportional to $X$.  More complex models may consider some convex loss function, assuming that 
less costly social distancing measures will be taken before more costly ones are.}

Figure \ref{fig:example_dynamics_and_cost} demonstrates the dynamics and outcomes of an outbreak under 
three different lockdown levels.  
Figure \ref{fig:example_dynamics} shows the ``usual'' graph of the number of infected people over time.    
With a lax $30\%$ lockdown, the disease spreads quickly and dies-out due to herd immunity 
only when over 80\%
of the population have been infected.  With a more strict
lockdown of $60\%$ we reduce $R_0$ to a somewhat lower value $R_e$, but still to a value above 1,
thus ``flattening the curve'' and reaching herd immunity at a lower infection percentage, leading to a total of around 40\% of the population being infected.
An even stricter lockdown of 80\% reduces $R_e$ below 1, stopping the spread of the
disease rather quickly.  Figure 1b depicts the trade-off between the mortality cost and the
economic cost of these three approaches.  With a 30\% lockdown we have a low
economic cost but a high mortality cost.  As we increase the lockdown rate, the economic cost rises,
while the mortality cost decreases.  However, as we continue to increase the lockdown rate, at 
a high enough rate, we manage to quickly eliminate the disease (the ``hammer'' \cite{Pueyo-hammer-and-dance}), 
allowing the economy to return to normal, resulting in low economic and low mortality costs. 

While a quick elimination of the disease is certainly desirable, if possible,
maintaining a strict enough lockdown rate is not always feasible in practice.  
Also, in reality, it is unlikely that the disease can be completely eradicated,
as a tiny number of new cases, e.g., coming from abroad, will likely keep trickling in.  
Thus we also consider an ``open system'' where a small flow of spontaneous infections occurs,
which makes eradication of the disease impossible and requires continued social distancing
measures, as is demonstrated in Figure \ref{fig:example_cost_map_with_flow}. 

In this paper we study how quarantines, testing, and tracking can help improve the basic trade-off between
mortality and economic costs beyond what is obtained from social distancing alone.

\begin{figure}[t!]
\centering
	\begin{subfigure}{.45\linewidth}
		\includegraphics[width=0.9\linewidth]{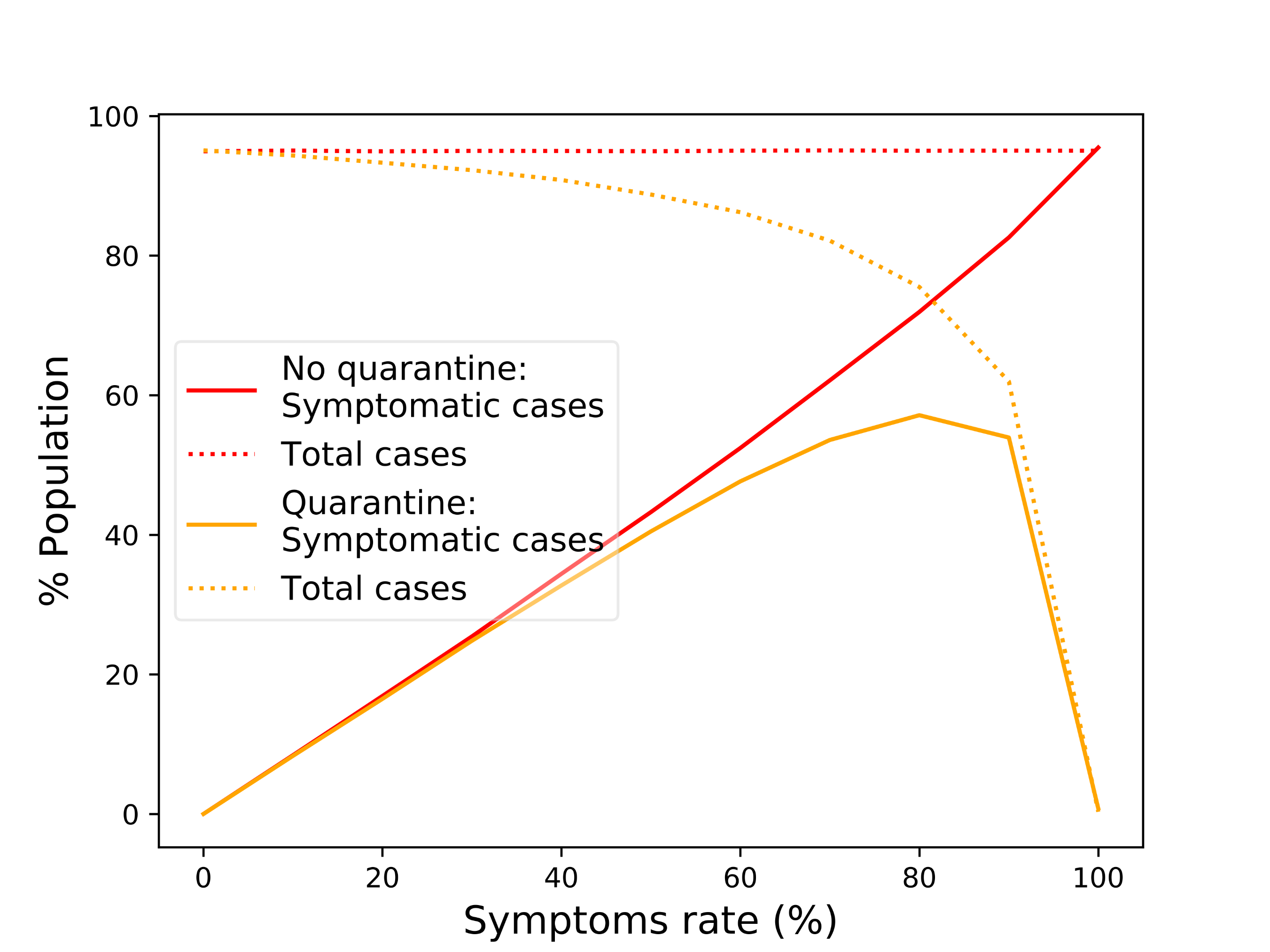}
		\caption{Total symptomatic cases}  
		\label{fig:symptoms-rate-infected}
	\end{subfigure}
	\begin{subfigure}{.45\linewidth} 
		\includegraphics[width=0.9\linewidth]{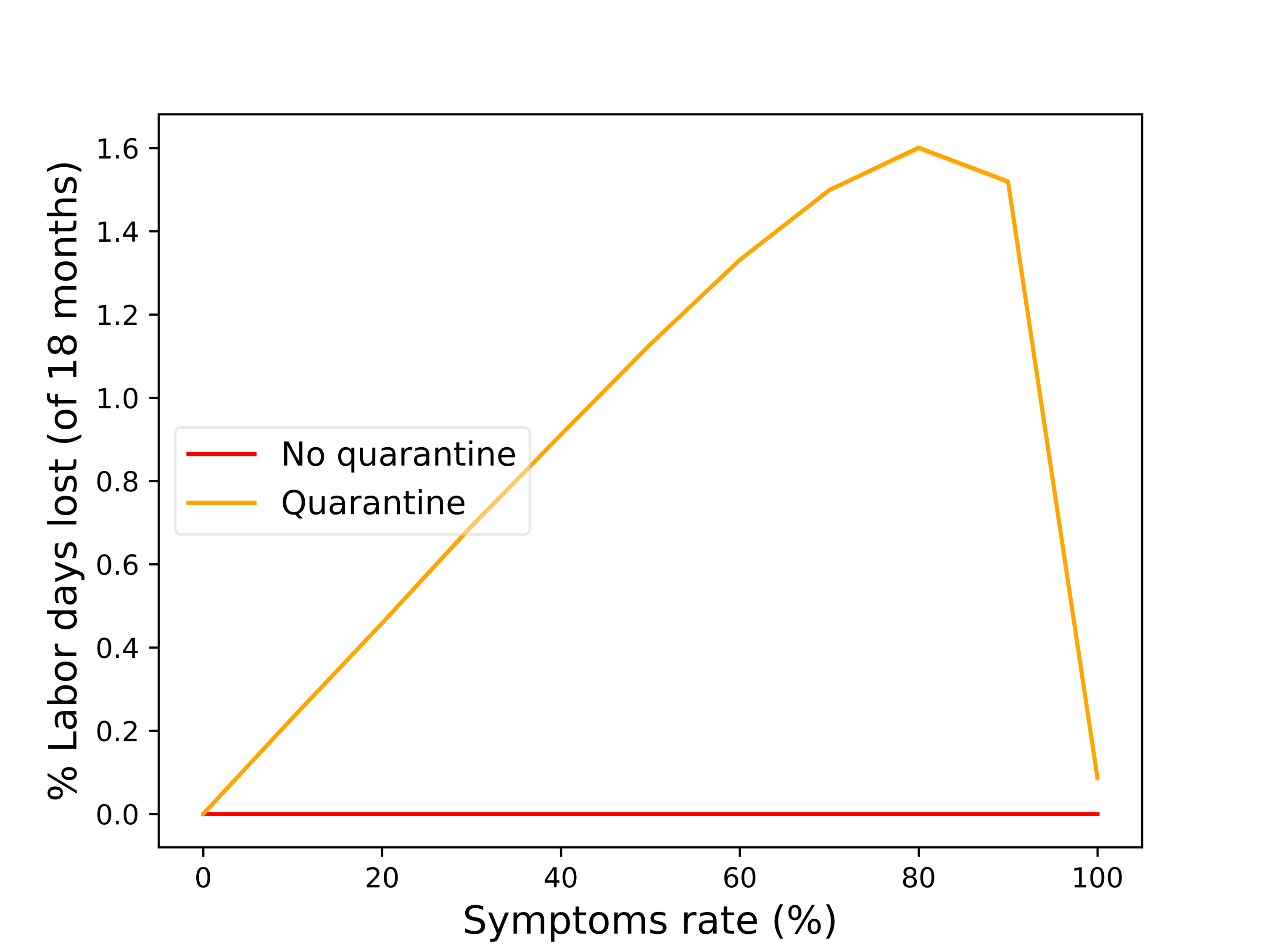}
		\caption{Economic costs}  
		\label{fig:symptoms-rate-cost}
	\end{subfigure}
	\caption{Effect of symptoms rate. \ref{fig:symptoms-rate-infected}: The full and dotted lines show the total population percentage of symptomatic and all infected patients, respectively, with and without quarantines, as a function of the symptoms rate. \ref{fig:symptoms-rate-cost}: Economic costs with and without quarantines as a function of the symptoms rate.}
	\label{fig:symptoms-rate}
\end{figure}

\subsection{Overview of Main Results}
\noindent
\textbf{Quarantine:} 
We start by looking at the effect of a basic quarantine: once an individual becomes symptomatic  
he is immediately quarantined until no longer infectious.  As expected,
and as shown in Section \ref{sec:quarantine}, quarantine reduces contagion and gives an effective increase in
the level of lockdown, without incurring the economic costs of tightening the lockdown.  
This is naturally especially significant near the critical rate of contagion $R_e=1$.  

A key parameter that determines the effect of quarantines is the fraction of asymptomatic patients.
If everyone is symptomatic then quarantine will completely eradicate the disease, while if
no one is symptomatic then in any case there will be no mortality. 
The difficult situation is where some fraction of cases are asymptomatic, in which case there is high mortality
but quarantine is not fully effective. This is demonstrated in Figure \ref{fig:symptoms-rate} that shows the total population percentage of symptomatic cases and the economic costs
as a function of the rate of symptoms among infection cases, with and without quarantine. 
The precise population mortality in every outbreak is some percentage of the total symptomatic cases, where the exact percentage depends on specific properties of the disease and of the population, on environmental conditions and conditions of the healthcare system. 
The figure shows that quarantines can stop the spreading of an infection only if almost all patients are symptomatic, while 
intermediate to high symptom rates of $50$-$90$ percents still lead to high mortality.

\vspace{3pt}
\noindent
\textbf{Tracking:} 
A promising intervention is to track previous contacts of any identified infectious person. People
who have been in contact with an infected person
need scrutiny for two possible reasons: on the one hand they may already be 
infected by this person without knowing it and, on the other hand,
they may be the asymptomatic source of this person's infection and may infect others.  
Obviously, tracking contacts of infected patients is a difficult and costly endeavor which seriously
infringes on the privacy of infected people and of their contacts.  Yet, in most western countries
citizens are required by law to cooperate with such an epidemiological investigation by
health authorities. In the current COVID-19 crisis, it has become clear that effective contact tracing at
scale requires automated tools such as cell-phone GPS-tracking which further heightens the 
privacy concerns. Trading-off the human-rights loss of privacy versus the
human-rights losses from lockdowns is a difficult social question, and requires a clear understanding
of the possible benefits of tracking. This is what we aim to study here.

\remove{
\nncomment{In our model, does contact tracing of a person ever find anyone that he has infected?  The
issue is that a person becomes infectious only when he is symptomatic at which point he is anyway quarantined and cannot infect anyone?}

\ykcomment{I think the answer is yes. For example if a person becomes symptomatic and so detected only, say, 3 days after he enters the infectious stage then he may have infected by that time his neighbors, who may have infect their neighbors.  We start the track and test recursion from the points that we know about, and I think there are (at least at the beginning) some disconnected ``clusters'' of infections that for some time we don't know about, they have a short lifetime before detected because some node in the cluster will eventually become symptomatic. }
}

\begin{figure}[t!]
\centering
	\begin{subfigure}{.45\linewidth}
		\includegraphics[width=0.9\linewidth]{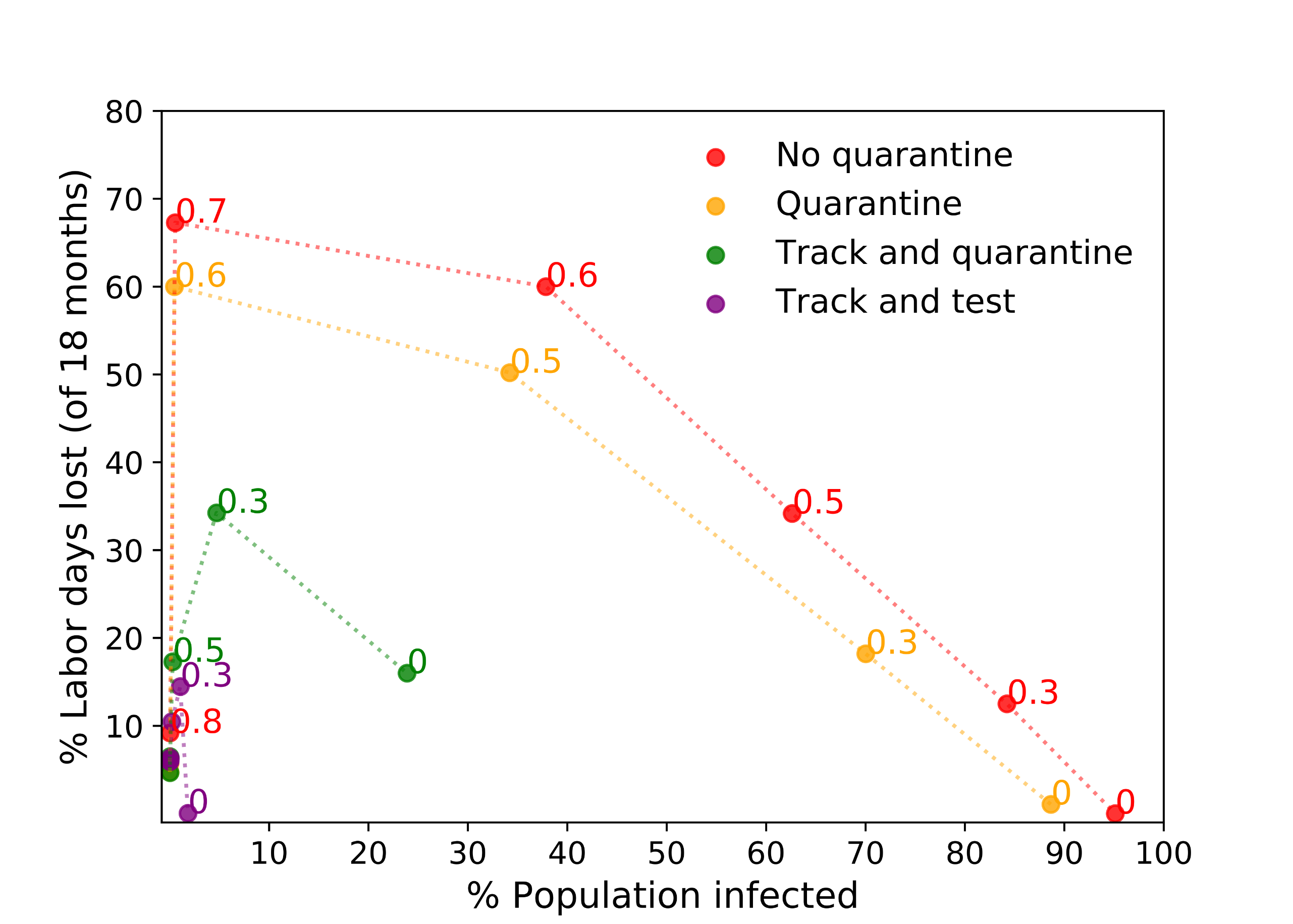}
		\caption{Closed system}  
		\label{fig:tracking_no_flow}
	\end{subfigure}
	\begin{subfigure}{.45\linewidth} 
		\includegraphics[width=0.9\linewidth]{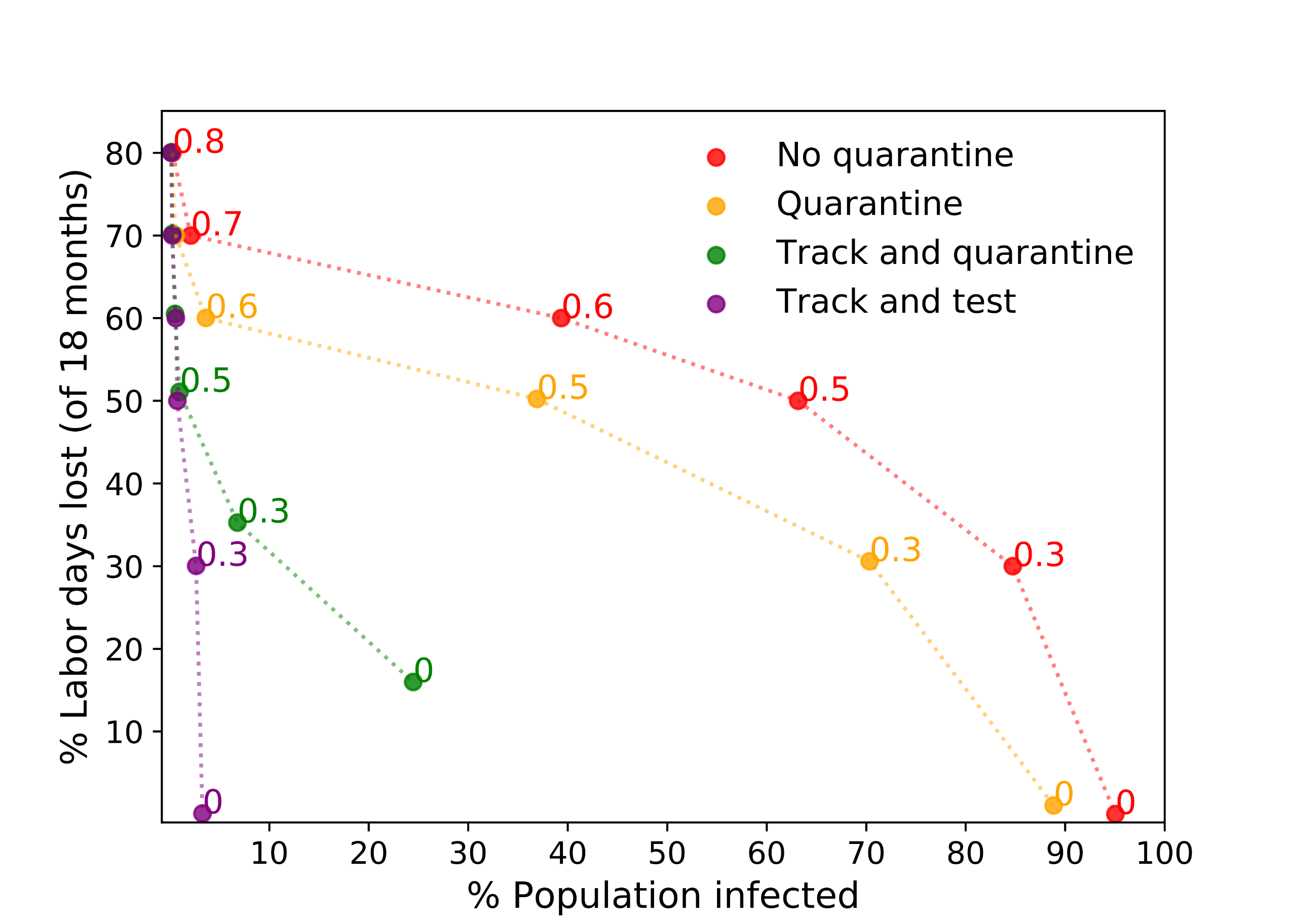}
		\caption{Open system}  
		\label{fig:tracking_with_flow}
	\end{subfigure}
	\caption{Tracking compared with the Quarantine method and the no-quarantine benchmark.}
	\label{fig:fixed_lockdown_tracking}
\end{figure}

\vspace{5pt}
\noindent
We consider two types of interventions after tracking a person's contacts: 
\begin{itemize}
\item {\em Track and Quarantine:}  This is the simple version of tracking: isolate infected 
individuals and all their recent direct contacts. 
\item {\em Track and Test:}  Test the direct contacts of an infected person and isolate only those
who have tested positive. Here, we naturally continue recursively with tracking and testing the contacts
of any person who tests positive. This method avoids the collateral damage of neighbor isolation: only infected individuals are put in quarantine. 
\end{itemize}

Figure \ref{fig:fixed_lockdown_tracking} compares the economic and mortality costs of these two interventions with the
basic Quarantine method and with the no-quarantine benchmark for a range of social distancing levels. These are shown both for a closed system
and for an open system with a small flow of incoming infections. 
The effectiveness of tracking is clearly visible at all levels of social distancing, 
with Track and Test being especially effective, often taking us below the critical threshold of contagion that eradicates the disease.   

We consider these simulation results to be a clear indication of the effectiveness of track-and-test strategies,
stressing the difficult normative question faced by democratic governments around the world
of whether and how to deploy contact tracing with its associated privacy loss.

\begin{figure}[t!]
\centering
	\begin{subfigure}{.45\linewidth}
		\includegraphics[width=0.9\linewidth]{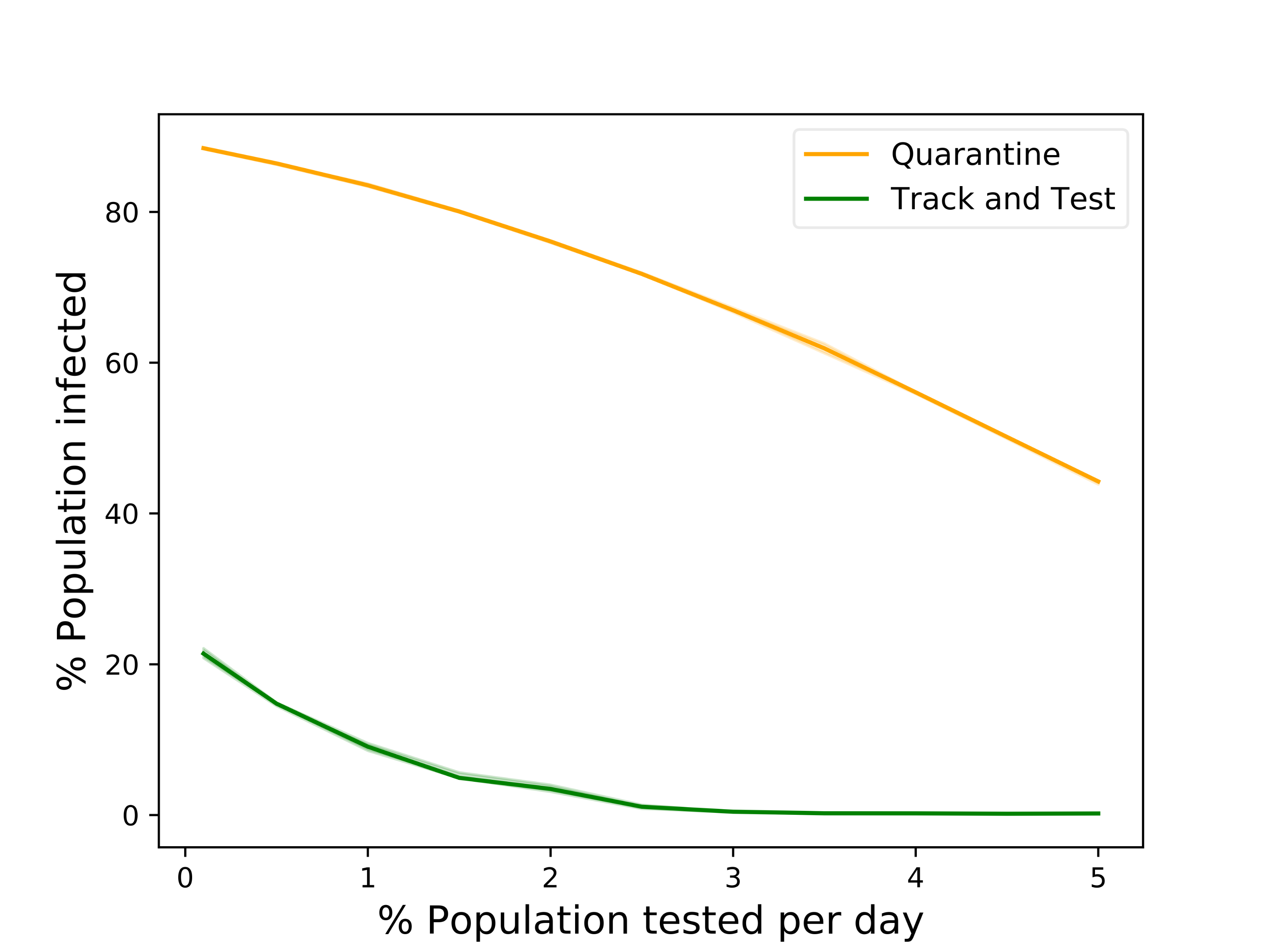}
		\caption{Total infected}  
		\label{fig:limited-tests-infected}
	\end{subfigure}
	\begin{subfigure}{.45\linewidth} 
		\includegraphics[width=0.9\linewidth]{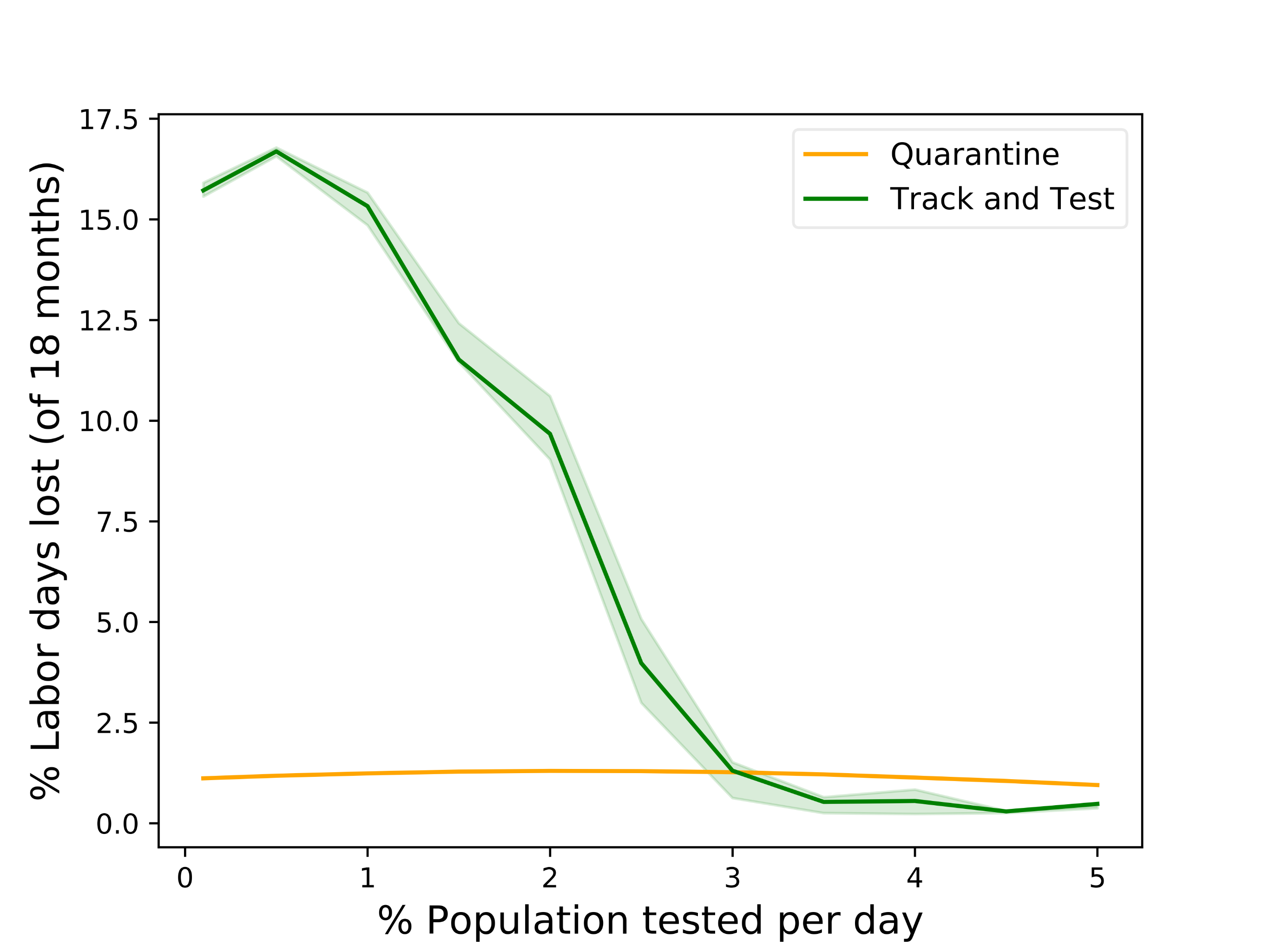}
		\caption{Economic costs}  
		\label{fig:limited-tests-cost}
	\end{subfigure}
	\caption{Effect of tests. A comparison between random tests and the Quarantine method and the bounded Track and Test method, as a function of the percentage of population tested each day.}
	\label{fig:limited-tests}
\end{figure}

\vspace{5pt}
\noindent
\textbf{Testing Capacity:} 
Our Track and Test method continued recursively with testing the contacts of
any infected person found until no more infections were detected. In the worst case this method 
requires an unbounded number of tests per day.
We also study a ``bounded-testing'' variant, where we only have a limited testing capacity,
and once this limit is reached we can no longer continue with testing.
Figure \ref{fig:limited-tests} shows the effect that the testing capacity has on the effectiveness of
the bounded, realistic, version of Track and Test 
in comparison with the Quarantine method combined with the same testing capacity where the available tests are administered at random.  
As Figure \ref{fig:limited-tests-infected} shows, while there is some advantage
to random tests and quarantine over straightforward quarantine without any testing, 
its effectiveness is a far cry from that of Track and Test. See more details in Section \ref{sec:limited-tests}.

\vspace{3pt}
\noindent
\textbf{On-Off Lockdown Policies:} 
Until now we have only considered constant lockdown policies. This is rather limiting in
our ``open system'' model that assumes a small steady flow of spontaneous infections, thus
ruling out the possibility of complete eradication of the disease. In such (realistic) models,
a natural strategy would be to dynamically relax the lockdown when there are only very 
few infected people, tightening it back when the number of infected people increases (this is the ``hammer-and-the-dance''
strategy popularized in \cite{Pueyo-hammer-and-dance}).  These dynamic lockdowns may be
able to out-perform static constant levels of lockdown.

\begin{figure}[t!]
\centering
	\begin{subfigure}{.45\linewidth}
		\includegraphics[width=0.9\linewidth]{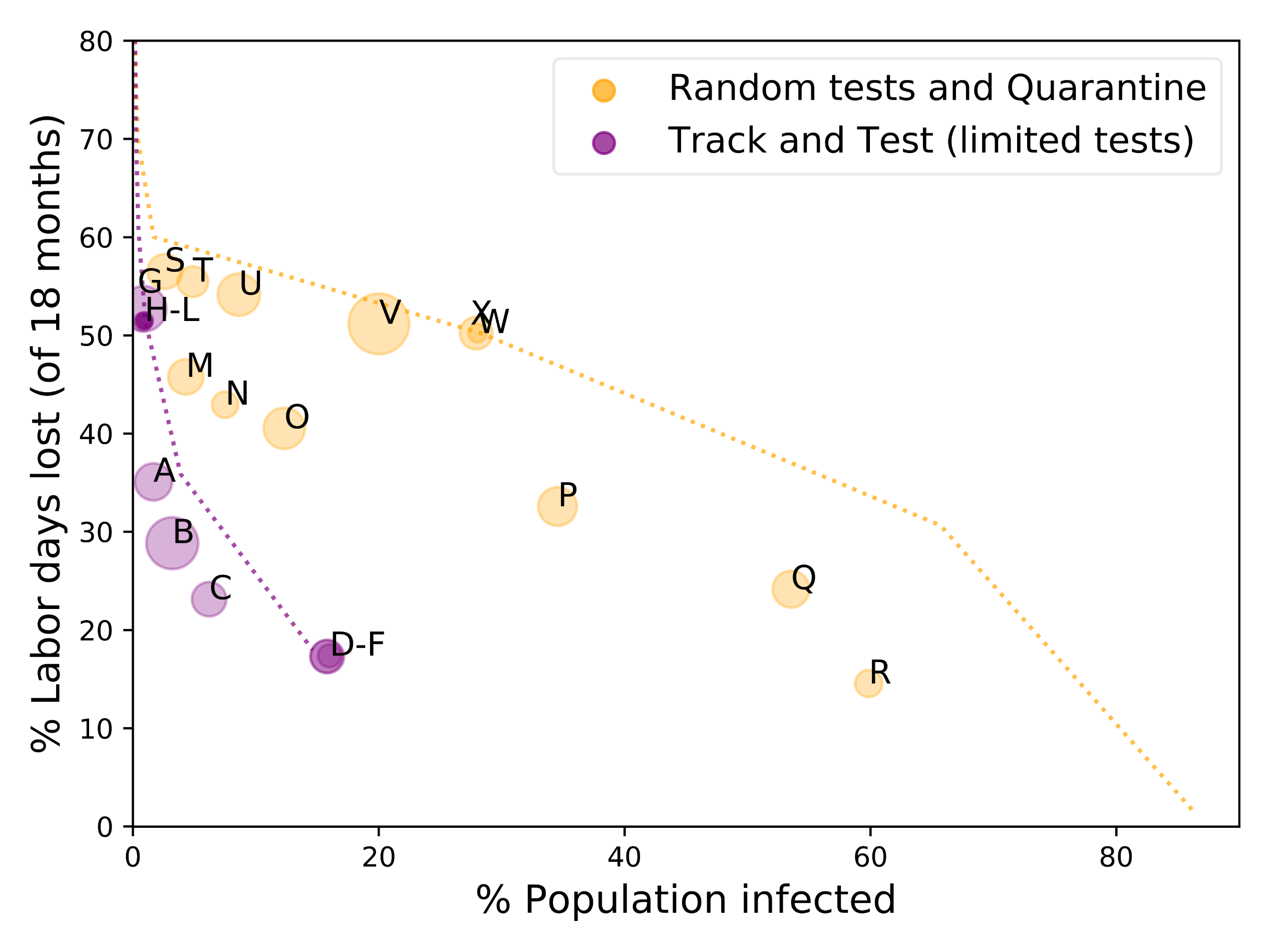}
		\caption{}  
		\label{fig:on-off-map}
	\end{subfigure}
	\begin{subfigure}{.45\linewidth}
	\vspace{-10pt}
		\includegraphics[width=0.935\linewidth]{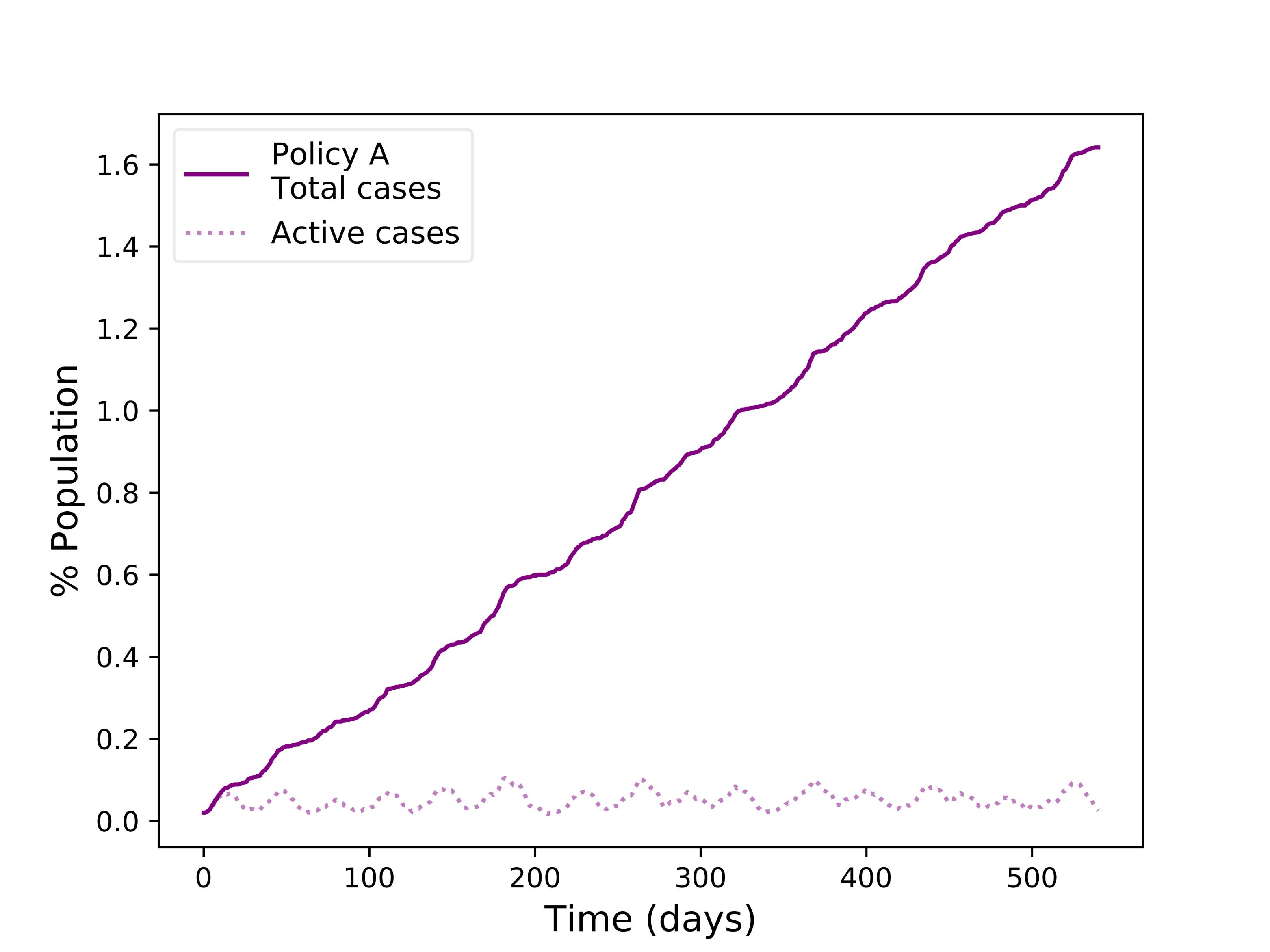}
		\caption{}  
		\label{fig:on-off-A-dynamics}
		\vspace{4pt}
	\end{subfigure}
\caption{On-Off policies. \ref{fig:on-off-map}: The cost map of the total percentage of labor days lost in $18$ months versus the total percentage of population infected. The letters near the markers indicate the different policies as shown in Table \ref{fig:table-On-Off}. Purple markers are policies with the bounded Track and Test method and orange markers are with the Quarantine method with random tests. The purple and orange dotted curves show a comparison to the results of fixed lockdown policies with bounded Track and Test and with Quarantine and random tests, respectively. \ref{fig:on-off-A-dynamics}: The dynamics of the outbreak under policy $A$.}
\label{fig:On-Off}
\end{figure}

Figure \ref{fig:on-off-map} depicts the cost map of percentage of labor days lost versus the total percentage of infected population for $24$ different On-Off policies. 
The two colors indicate whether the lockdown policy is applied with the limited tests version of the Track and Test method or with random tests and the Quarantine method; the population percentage of tests per day for both methods is $0.5\%$. 
Table \ref{fig:table-On-Off} shows the details of the algorithms corresponding to the policies in the figure. The 12 left columns are policies with the Track and Test method and the right columns are with random tests and Quarantine. 
The dotted curves show the comparison to the costs of using fixed lockdowns with the two methods and the same number of tests.
It can be seen that in the presence of incoming infections, dynamic lockdown policies are more efficient than fixed lockdowns 
both for Quarantine with random tests and for the Track and Test method. 
Also in this setting with a limited testing capacity and in combination with dynamic lockdowns, Track and Test policies show a clear advantage over Quarantine policies. See Section \ref{sec:On-Off} for further analysis of the different dynamic lockdown policies. 

One policy that stands out in Figure \ref{fig:on-off-map} is policy $A$. Policy $A$ combines the Track and Test method with a strong lockdown of $80\%$ and a full removal of the lockdown once the known number of active cases becomes very low -- below $0.01\%$ of the population. The lockdown is ``re-activated'' once this number is doubled. 
Figure \ref{fig:on-off-A-dynamics} shows the dynamics of the spread of the infection under policy $A$. It can be seen that this type of low-threshold dynamic lockdown policy in combination with tracking and a limited number of tests manages to restrict the spread of the infection to below $2\%$ of the population in a period of $18$ months at significantly lower costs than the non-tracking or the fixed-lockdown alternatives.

\section{Preliminaries}

\subsection{Background}
The early roots of mathematical modeling of the spread of epidemics and the effect of interventions trace back to the classic work of Bernoulli from 1776 (see \cite{bernoulli-1776-a,bernoulli-1776-b}), where he used differential equations to analyze the effect of inoculation as an intervention tool for reducing mortality in smallpox outbreaks. About a century and a half later, the modern SIR model was analyzed in the seminal work of Kermack and McKendrick in 1927 \cite{kermack1927}, based on an earlier model of Ross and Hudson \cite{ross1917}. The SIR model is based on three coupled differential equations that describe the dynamics of the sizes of three populations: Susceptible, Infected and Removed. The SEIR model is an extension that adds the Exposed state, which is a non-infectious incubation stage. 
The steady-state result of the total spread of an infection is the same for the two models. 

The original model of Kermack and McKendrick is deterministic and considers the population as a homogeneous continuum. A natural extension of the model is to a Markov chain model with discrete ``compartments'' that count the population in each of the states. Stochastic  compartmental models of this type allow to simulate more complex settings where the population is no longer homogeneous but is divided to different groups (such as age groups) that have different transition rates between states, or to add more states such as hospitalization states and different disease severity levels; of course this flexibility comes with the cost of more free parameters in the models. Several such models were recently designed for simulating outbreaks in the current COVID-19 crisis with different interventions, some of which are used by advisory teems for policy makers or as publicly available open source tools (e.g., \cite{walker2020report, coexist}). 
A basic characteristic of compartmental models is that they count the population size in each state and the flows of populations between the different states but, importantly, they do not have identities for individuals and do not model individual interactions. 

SIR and SEIR models have also been studied on networks. 
In the network-based model, individuals can be identified and their interactions can be tracked; this is required in order to directly study the effect of tracking,  
whereas population-based models require to a priory abstract and model the effects of tracking or quarantines \cite{Science-ferretti2020quantifying}. 
Previous work on SIR and SEIR models on networks focused mainly on basic statistical properties of these models on several network types \cite{newman2002spread,liu2017analysis}, or on the effect of targeted vaccination \cite{Nature2004Networks,altarelli2014containing}. The current work focuses on the effect of tracking, quarantines and tests on SEIR models with individual interactions in networks.

\subsection{Model and Setting}\label{sec:model}
We now give the exact details of our model and in the following section discuss the outcomes of different mitigation policies under this model. The model is a direct adaptation of the SEIR model to dynamic networks. 
We augment the SEIR model with one additional parameter that describes the average fraction of symptomatic cases among infected individuals. As discussed in the introduction, the symptomatic cases rate is a crucial property of an infection and its addition to the model is necessary for evaluating the effects of tracking and quarantines.

\subsubsection{Network Interaction SEIR Model}
We adhere in our modeling to the basic SEIR model and use a uniform infection probability $p$ per interaction and uniform disease progression times. The infection probability $p$ is related to the basic reproductive number $R_0$ 
in the SEIR model as follows. Let $m$ be the average number of interactions of an infected individual during the whole infectious stage, then $R_0 = m \cdot p$ is the average number of infections caused by this individual when the rest of the population is in the susceptible state. Throughout the paper we use $R_0 = 3.6$.\footnote{$R_0=3.6$ is plausible for the COVID-19 \cite{liu2020reproductive,flaxman2020report}; other values simply correspond to lockdown level shifts.}
The infection rate and the time scales in the experiments are chosen on the one hand to be in the correct realistic range of epidemics of the type of the COVID-19, but on the other hand, we do not aim to give a detailed modeling of a specific outbreak, but rather to study and demonstrate basic effects and basic trade-offs of intervention approaches. We thus avoid modeling details such as age or the health conditions of individuals other than that which is related to the infection.  

Once an individual interacts with an infectious person there is a probability $p$ that he becomes infected and moves to the exposed state. The progression of the disease is then governed by two time constants and an additional parameter to describe symptoms. Let $\tau_{{\scriptscriptstyle E}}$ be the average duration of the incubation time of the disease (the exposed state) and let $\tau_{{\scriptscriptstyle I}}$ be the average duration of the infectious state. After this time the patient moves to a removed state (either recovers, is isolated until no longer infectious, or with some probability dies). We use the time constants $\tau_{{\scriptscriptstyle E}} = 6$ days and $\tau_{{\scriptscriptstyle I}} = 8$ days. 
Let $q$ be the symptoms rate, which is the average fraction of infected individuals who develop symptoms at some time during their infectious stage. We assume a constant probability per day to develop symptoms at each day of the infectious stage, such that the total probability for each individual to develop symptoms at some time during the whole infectious stage equals $q$. This allows to model both asymptomatic and pre-symptomatic infection cases. In the experiments presented in the paper we use $q=0.5$.\footnote{For the COVID-19, symptoms-rate estimates range between $0.2$ and $0.9$ by different sources \cite{mizumoto2020estimating,nishiura2020estimation,day2020covid,CEBMsymptomatic}.} We also study the effect of different values of this parameter (see Section \ref{sec:quarantine}). 

We consider the dynamics of the spread of an infection in discrete time, in steps of one day. The results of the experiments are measured across a time span of $540$ days (roughly a year and a half). 
A population of $N=10^5$ individuals is represented in the model as a set $V_t$ of vertices in an interaction graph.  
Let $G_t = \left\{V_t,E_t\right\}$ be the interaction graph at time $t$. 
Each vertex has a unique identity and a private internal state at time $t$ that includes the following properties: 
(1) Infection state: susceptible, exposed, infectious or removed; (2) Symptoms state: symptomatic or not symptomatic. Only individuals in the infectious state can become symptomatic; (3) Quarantine state: currently quarantined or not quarantined, and the time since the beginning of the quarantine. Quarantines are for a period of 14 days.
An (undirected) edge $(i,j)_t$ in this graph means that the two individuals represented by vertices $i$ and $j$ interacted at day $t$.

At each day $t$ a new subset of the possible edges describes the interactions of this day. 
In principle, any time series $(E_1,...,E_t)$ of interactions can be used with the model. The basic network model we use is of uniform mixing, which is similar to the SEIR model. In the appendix we show qualitatively similar results also for geographic small world networks \cite{kleinberg2000small}, where the vertices are spatially located in two dimensions and have frequent interactions with their close relations, but also have a small number of long range interactions with a probability that diminishes quadratically with the geographic distance of the interaction. 

To compare between mortality costs under different intervention methods we measure the total size of the outbreak, which is the eventual population percentage of infected individuals. 
Of course, various other 
models can be designed for mortality prediction in specific settings, where each such model results in some range of mortality percentages of the total size of the outbreak. Precise mortality prediction requires many details such as hospitalization times, populations at risk, intensive care units capacity, availability of medical equipment and staff and other parameters, and necessitates additional modeling assumptions. 
The total size of the outbreak is a fundamental quantity of the dynamics and thus it is less prone to modeling errors, and provides sufficient and, in our view, more reliable information for distinguishing between different mitigation methods.

\subsubsection{Intervention Policies}\label{sec:intervention-policies}
\noindent
\textbf{Lockdowns:} We use a single lockdown level parameter to model the effect of all social distancing measures. For example, a lockdown level of $0.5$ (or $50\%$) means that after the edges, $E_t$, for a new day are determined, each of these edges will be deleted with probability $0.5$. For the economic cost of lockdowns we use a linear model that assumes, e.g., that in a lockdown of $50\%$, half of the labor force is at work. As mentioned in the introduction, one may also assume more complex models of convex costs that take into account that less costly measures of social distancing are taken first, or to consider a non-linear dependence of the costs of lockdowns on their duration. 

\vspace{3pt}
\noindent
\textbf{Quarantines:} Quarantined individuals are assumed to have no infectious interactions for a period of $14$ days. In the simulations, after the set of edges for a new day is determined, and before the actual interactions are simulated, all the edges that are connected to quarantined individuals are removed.

\vspace{3pt}
\noindent
\textbf{Tracking:} Let $H_t =\left\{V_t, \bigcup\nolimits_{t'=t-T + 1}^t E_{t'}\right\}$ be the tracking network at time $t$, where the tracking time $T$ determines how many days of interactions are being tracked. We use a tracking time of $10$ days, which is a little above  the duration of the infectious stage. 
The tracking network $H_t$ identifies the individuals in the population as vertices and its edges show all the past interactions of each individual in a recent time window of $T$ days. 
The health authorities can identify the vertices in $H_t$ and observe its edges at the end of each day, and then decide on a policy for day $t+1$. The internal properties of the vertices are not observed, except for these that are related to quarantines. Regarding the infection state, only a coarser level of information can be detected: (i) Symptomatic individuals are reported as infected; (ii) Individuals that are tested positive are also reported as infected, but without the specific knowledge of the exact states of their disease. In Section \ref{sec:tracking} we discuss different methods in which the tracking network can be used as a mitigation tool. 


\section{Results}\label{sec:results}

In this section we discuss the effects of the intervention tools presented in Section \ref{sec:intervention-policies} -- lockdowns, quarantines, tracking and testing -- on the dynamics and outcomes of an epidemic in the SEIR model in dynamic networks of interacting individuals. 
We first discuss the effects of lockdowns and of quarantine of infected individuals on the economic cost and on the total spread of the infection. Then, we evaluate the potential benefits of tracking the contacts of infected individuals and the added value of tests. Finally,  we discuss dynamic lockdown policies and how do they interact with tracking and testing  and with quarantines.   

\subsection{Lockdowns and Quarantine}\label{sec:quarantine}
Lockdowns are a direct tool for limiting the progression rate of an infection: a ``blanket lockdown'' that reduces on average the probability of each interaction by a fraction $X$ leads to an immediate reduction in the infection rate, reducing the basic reproductive number from $R_0$ to $R_e = (1-X)R_0$. In principle, a strict lockdown for a sufficient period of time can eradicate an infection completely. However, lockdowns have two key caveats as mitigation or containment tools: the first caveat is the high economic and social costs of lockdowns, and the second is that a full extinction state of the infection is not always feasible in practice due to its sensitivity to new incoming infections.

In Figure  \ref{fig:fixed_lockdown_tracking}, the red curves show the effect of using only lockdowns as a mitigation tool. 
The figures depict the cost maps of the total percentages of infected population and the percentages of labor days lost throughout a period of $18$ months, in a closed system (Figure \ref{fig:tracking_no_flow}) and with a small flow of incoming infections (Figure \ref{fig:tracking_with_flow}). The numbers near each marker indicate the lockdown level and the dotted lines are linear interpolations. 
The results show a qualitative difference between closed and open systems: 
In a closed system, as seen in Figure \ref{fig:tracking_no_flow}, it is possible to hold a strong lockdown of $80\%$ that reduces the effective reproductive number to below $1$ and to eradicate the infection completely. This leads to an overall  low percentage of infected population and to relatively low costs, since the lockdown can be removed after a short period  once there are no more infections. However, Figure \ref{fig:tracking_with_flow} shows how adding even a small incoming flow of infections (a single new infection per week in a 100,000 population) leads to a different result; when relying solely on lockdowns, heard immunity is not reached and, in the lack of additional tools to restrain the spread of the infection, the lockdown cannot be removed and the economic cost remains high. The curves show that in both closed and open systems there is a trade-off between mortality and labor days lost, but in systems that are $100\%$ closed this trade-off breaks when moving from a subcritical to a supercritical lockdown level, while in the presence of some leakage of new infections the trade-off persists.

  
Next, we turn to look at the effect of quarantines. The yellow curves in Figure \ref{fig:fixed_lockdown_tracking} show the costs when symptomatic infected individuals are put in quarantine (see details of the Quarantine method in Section \ref{sec:model}). It is clearly seen that the Quarantine method dominates the no-quarantine benchmark. I.e., for every fixed cost, Quarantine leads to a lower total infection percentage, and for every fixed infection percentage Quarantine has a lower cost than the no-quarantine benchmark. 
When looking at fixed lockdown levels and comparing the no-quarantine (red) and Quarantine (yellow) methods, 
we see that at low lockdown levels, quarantining symptomatic cases creates some trade-off between economic cost and the mortality rate, mainly in the closed system condition. This happens because a significant part of the population are symptomatic and quarantined, leading to a non-negligible loss of labor days. However, this trade-off breaks when the lockdown level is higher and closer to the critical value. E.g., at lockdown levels of $0.6$ in both closed and open systems, Quarantine has lower economic costs and lower mortality rates than the no-quarantine benchmark. 

The effectiveness of quarantines depends on the 
rate at which infected individuals develop symptoms. 
In an extreme case where all infected individuals are symptomatic and detected at an early time of their infectious stage, quarantining infected individuals may be sufficient to stop an outbreak.  
On the other extreme, if there are almost no symptomatic cases, then the infection is not considered dangerous and even a widespread outbreak is not a serious concern. The situation of intermediate values is harder to control: When only part of the infected individuals develop symptoms, and symptoms may start either early or late during the infectious stage of the disease, asymptomatic and pre-symptomatic individuals act as ``invisible spreaders'' of the infection, leading to high mortality. Figure \ref{fig:symptoms-rate} shows the total population percentage of symptomatic cases as a function of the rate of symptoms, with and without quarantines. 
As can be seen, a high symptoms rate exerts ``force'' in two opposite directions: On the one hand, symptomatic cases can be detected and isolated, but on the other hand, a higher rate of symptoms naturally leads to more symptomatic cases. The tension between these two forces is expressed in the non-monotonic cost curves, both having a maximum point, with intermediate symptom rates resulting in high costs.

\subsection{Tracking}\label{sec:tracking}

\begin{figure}[t!]
\centering
	\begin{subfigure}{.45\linewidth}
		\includegraphics[width=0.88\linewidth]{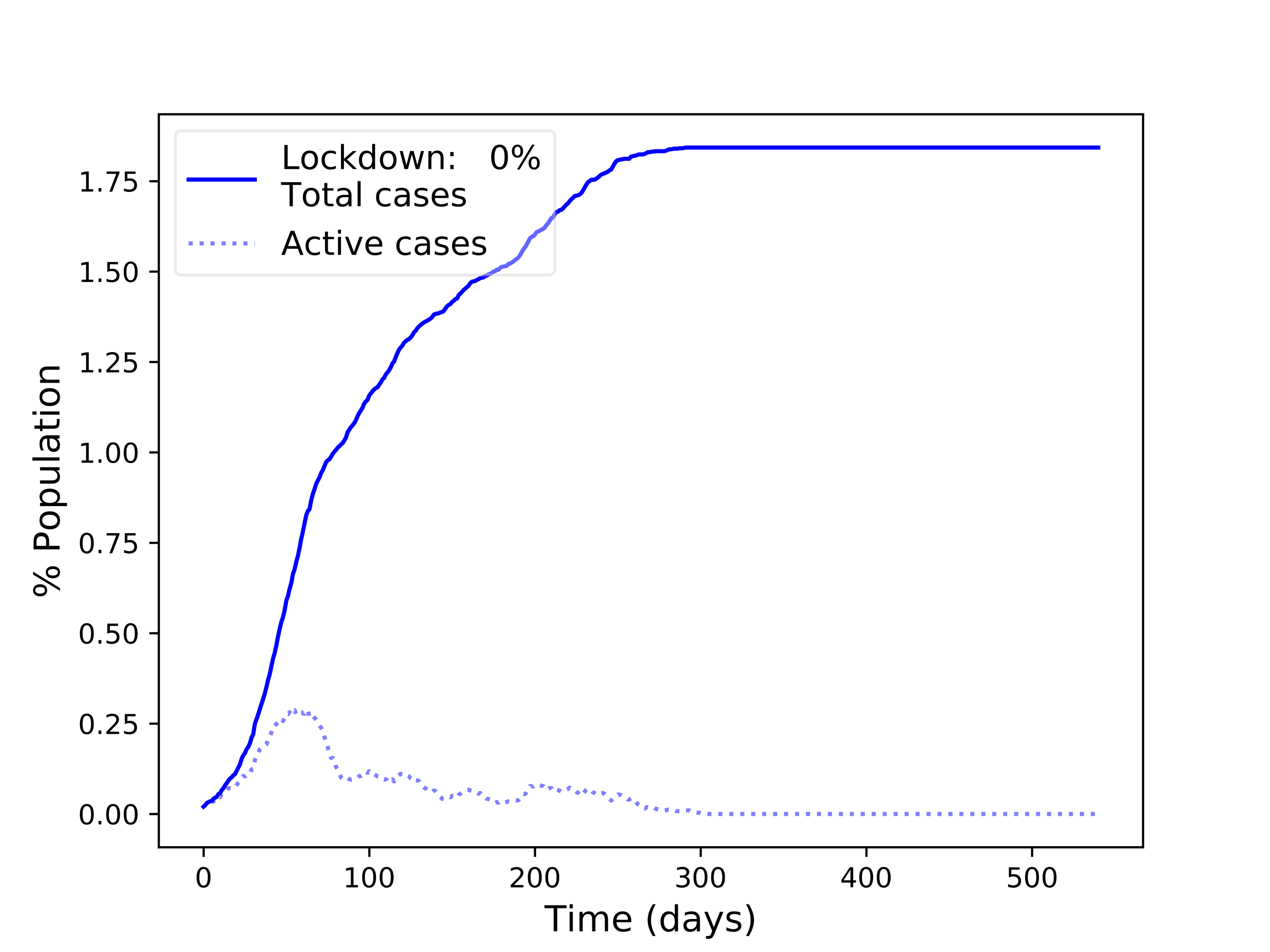}
		\caption{Closed system}  
		\label{fig:track-and-test-dynamics-closed}
	\end{subfigure}
	\begin{subfigure}{.45\linewidth} 
		\includegraphics[width=0.9\linewidth]{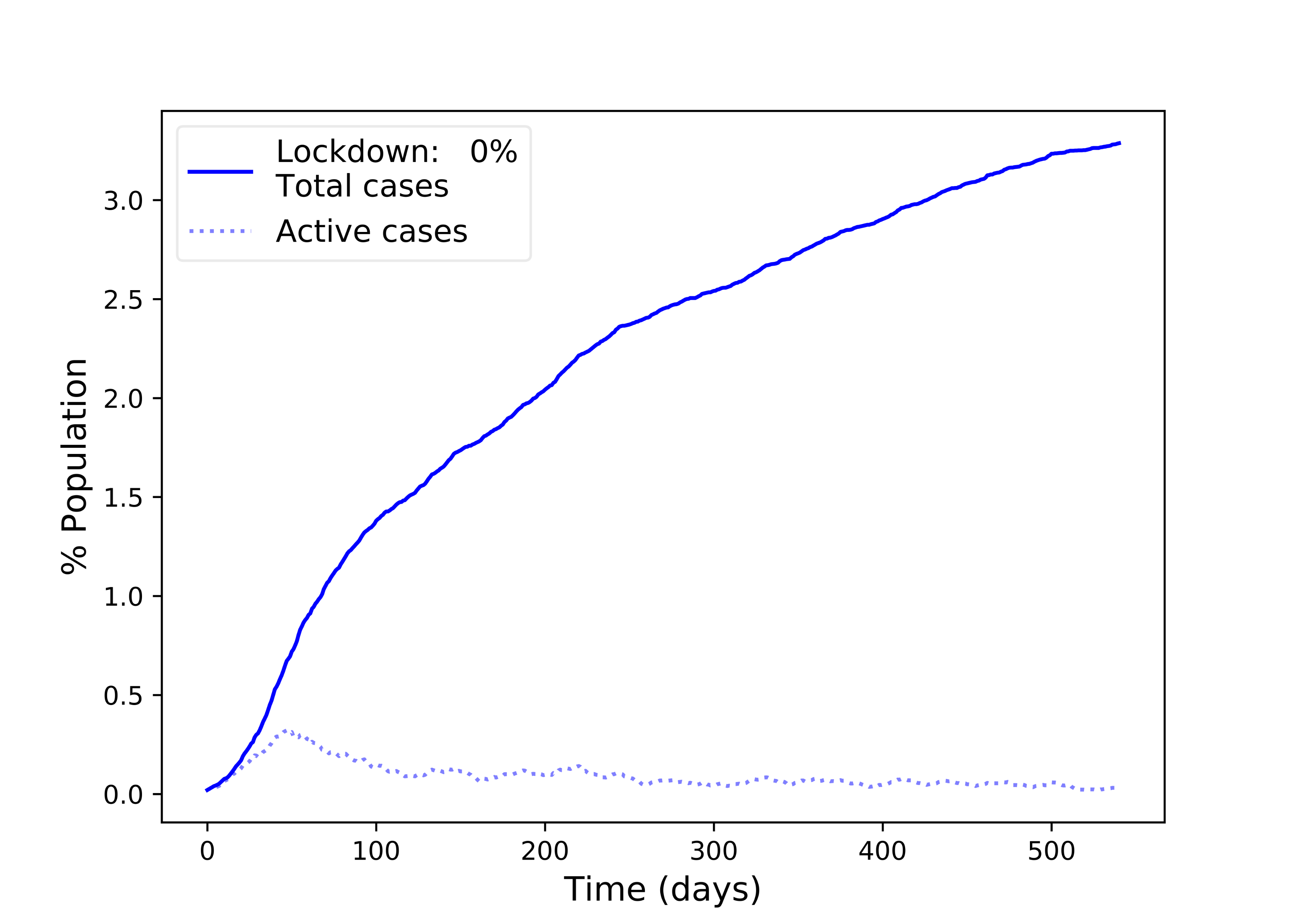}
		\caption{Open system}  
		\label{fig:track-and-test-dynamics-open}
	\end{subfigure}
	\caption{Track and Test without any lockdown: Outbreak dynamics. }
	\label{fig:track-and-test-dynamics}
\end{figure}

Next we describe the results of the tracking methods described in Section \ref{sec:intervention-policies} and compare these results to the basic Quarantine method. The tracking methods we compare are Track and Quarantine that quarantines symptomatic individuals as well as everyone they interacted with in the past 10 days, and the Track and Test method that recursively performs tests based on direct interactions with symptomatic individuals in the past 10 days and quarantines at every day only the detected infected individuals.  

The green and purple curves in Figure \ref{fig:fixed_lockdown_tracking} show the cost map of the Track and Quarantine method and the Track and Test method, respectively. A comparison with the results of the basic Quarantine method (in yellow) shows a clear-cut advantage of the two tracking methods over the Quarantine method in terms of mortality under all lockdown levels, in both open and closed systems. 
In terms of economic costs, at low lockdown levels of $30\%$ or below, the Track and Quarantine method quarantines a large fraction of the population due to their interactions with infected individuals and has the highest cost. In contrast, at higher lockdown levels Track and Quarantine manages to contain the infection and its economic costs drop and become lower than these of the Quarantine method. 
The Track and Test method that quarantines only infected individuals leads to low infection spread and economic costs, outperforming the other methods.

In Figure \ref{fig:fixed_lockdown_tracking} it is particularly interesting to notice the zero lockdown points of 
	the Track and Test method. The figure shows that both in open and closed systems, the Track and Test method manages to suppress the spread of the infection and practically achieves containment of the outbreak without using any lockdown. 
Figure \ref{fig:track-and-test-dynamics} shows the dynamics of the infection under the Track and Test method without any lockdown in open and closed systems. In the closed system (Figure \ref{fig:track-and-test-dynamics-closed}), Track and Test eliminates the infection to zero after about a year, with a total of less than $2\%$ infected population. With an incoming flow of infections (Figure \ref{fig:track-and-test-dynamics-open}) complete elimination is impossible, but here the spread of the infection under the Track and Test method is slow and sub-linear, reaching only less than $4\%$ of the population in $18$ months.    

\begin{figure}[t!]
\centering
	\begin{subfigure}{.45\linewidth}
		\includegraphics[width=0.9\linewidth]{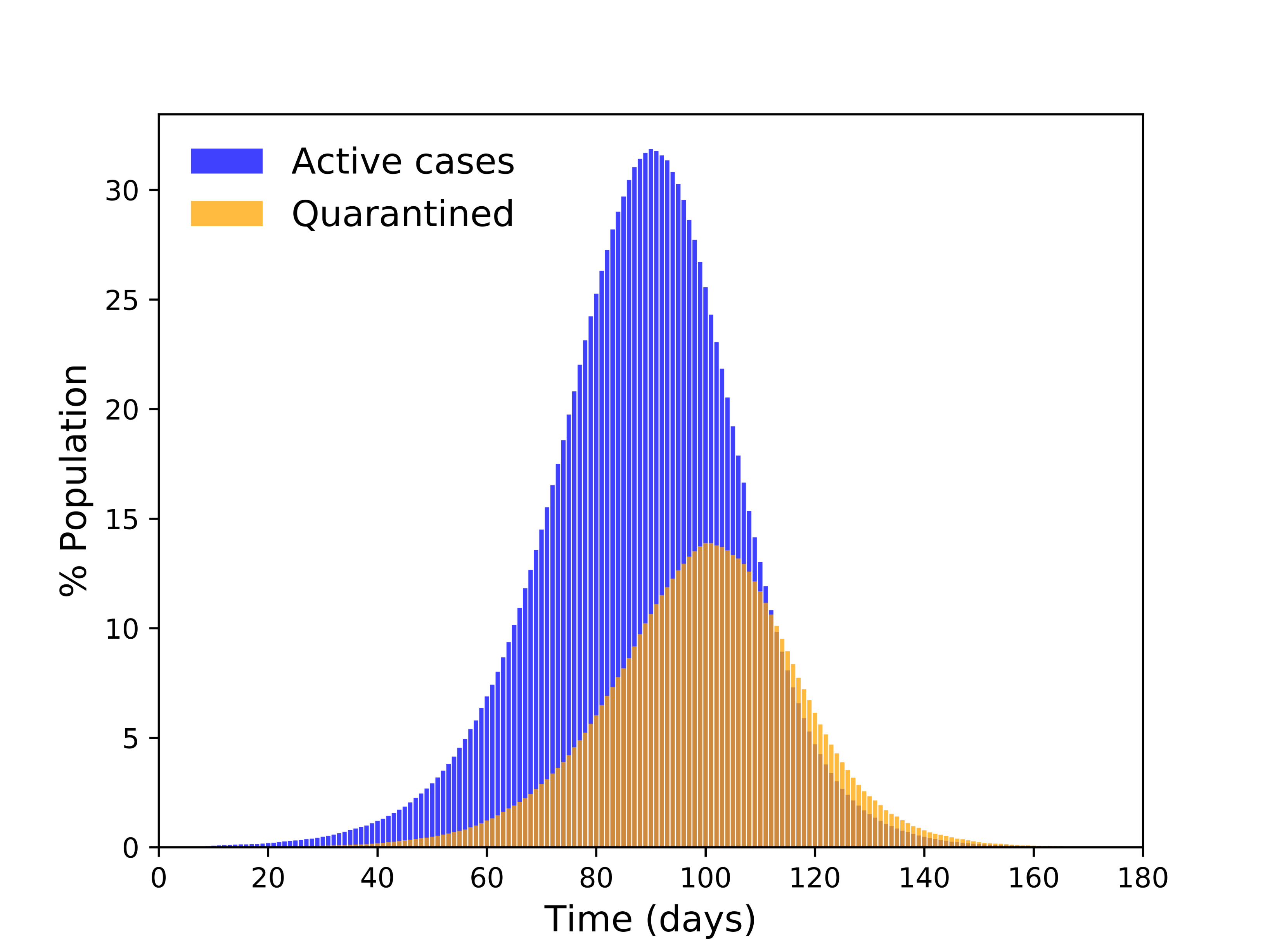}
		\caption{Quarantine}  
		\label{fig:active_and_quarantined_a}
	\end{subfigure}
	\begin{subfigure}{.45\linewidth} 
		\includegraphics[width=0.9\linewidth]{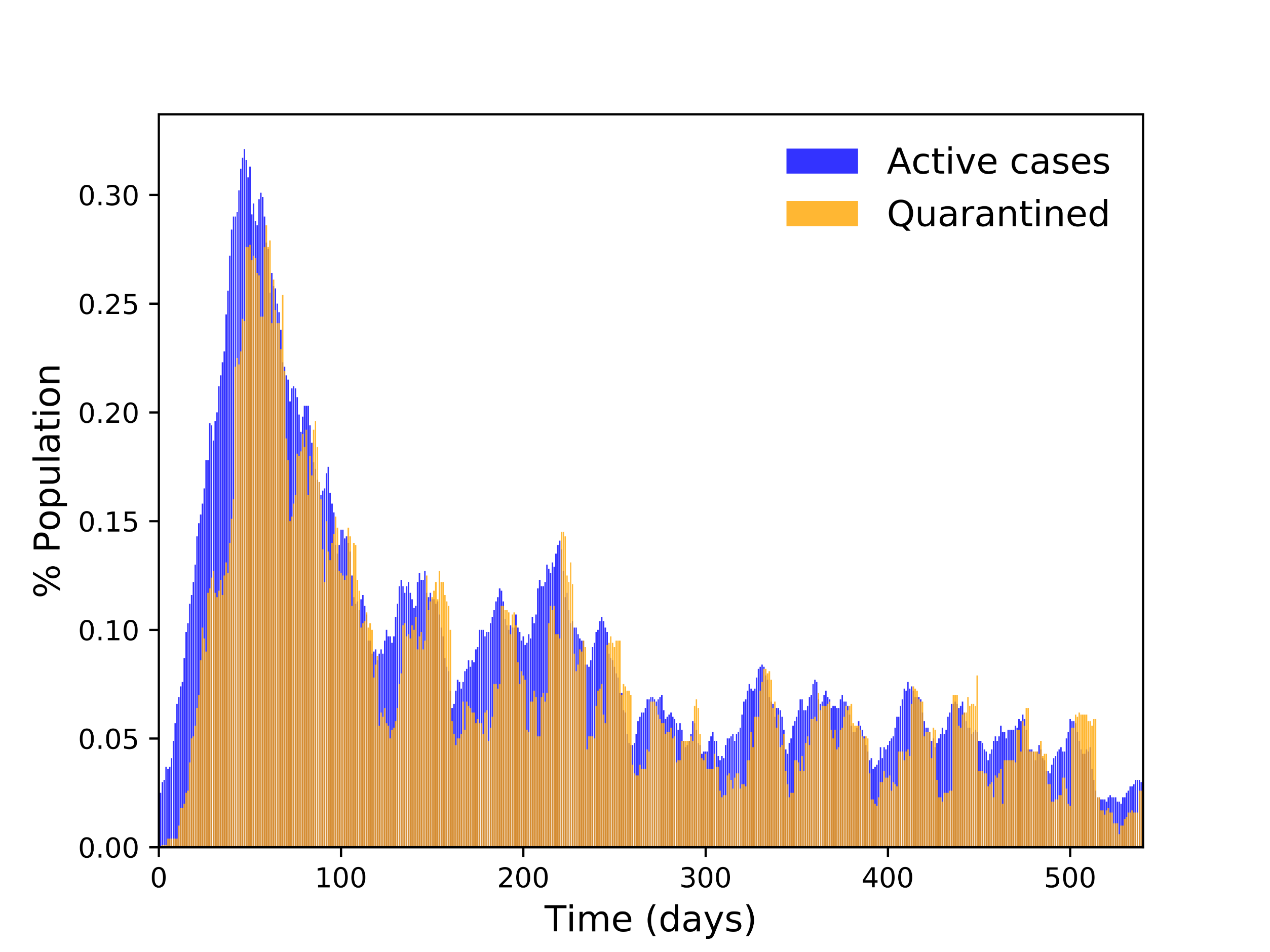}
		\caption{Track and Test}  
		\label{fig:active_and_quarantined_b}
	\end{subfigure}
	\caption{Active and quarantined cases.}
	\label{fig:active_and_quarantined}
\end{figure}

To shed more light on the process in which the Track and Test method manages to outperform the basic Quarantine method and suppress the outbreak we look at the dynamics of the number of actual active cases (exposed and infectious states) and the count of currently quarantined individuals for the two methods. Notice that the actual active cases count does not include recovered cases while the quarantined cases typically do include some recently recovered patients.\footnote{Adding daily tests to quarantined patients can remove this inefficiency of quarantines,  however, if the testing capacity is limited it is not clear if this practice is beneficial. In our model quarantines are for a fixed period.} 
Figure \ref{fig:active_and_quarantined_a} shows the dynamics of the active cases count across time (in blue), presented as percentages of the population, and the dynamics of the percentage of the population currently in quarantine (in orange) for the Quarantine method. Figure \ref{fig:active_and_quarantined_b} shows a similar plot for the Track and Test method. Both figures are presented in the more realistic open-system condition. 
%
We observe two problems in the Quarantine method: First, the number of quarantined cases reaches significantly lower levels than the number of actual cases, and second, Quarantine is ``lagging behind'' the dynamics of the actual infected cases. 
Both of these problems stem from the fact that the Quarantine method relies on passive detection of symptomatic cases; these cases consist only a part of the actual cases due to a large fraction of asymptomatic cases and are detected with a latency due to the incubation and pre-symptomatic time of the disease. 
The Track and Test method manages to abridge both of these difficulties, as it does not rely only on natural detection of symptomatic cases. Instead, these cases are used as seeds to track and then quarantine infection chains, quarantining also exposed and asymptomatic  individuals. Figure \ref{fig:active_and_quarantined_b} shows that the quarantined cases reach close to the actual active cases and very little lag effect can be seen. 
The advantage of tracking and testing is seen in the two orders of magnitude difference in scale between the two figures ($30\%$ vs. $0.3\%$).

\subsection{Limited Testing Capacity}\label{sec:limited-tests}
We have seen that the Track and Test method outperforms the basic Quarantine method by a large gap in both the scale of spread of the infection (and thus the mortality rate), and in the economic costs of the mitigation efforts. An issue that might arise in the context of the Track and Test method is that in the worst case, this method requires an unbounded number of tests (limited only by the size of the population). It could be that Track and Test simply tests the whole population in a short period of time; such a strategy, if it were possible to implement, would stop the infection, and it does not require any tracking. For (an extreme) example, if the whole population was to be tested in a single day, all infected individuals could be quarantined and the infection would not spread. To test whether this is the case we first look at the actual number of tests run by the (unbounded) Track and Test method, and then look at a bounded-tests variant of the method.

\begin{figure}[t!]
\centering
	\begin{subfigure}{.45\linewidth}
		\includegraphics[width=0.9\linewidth]{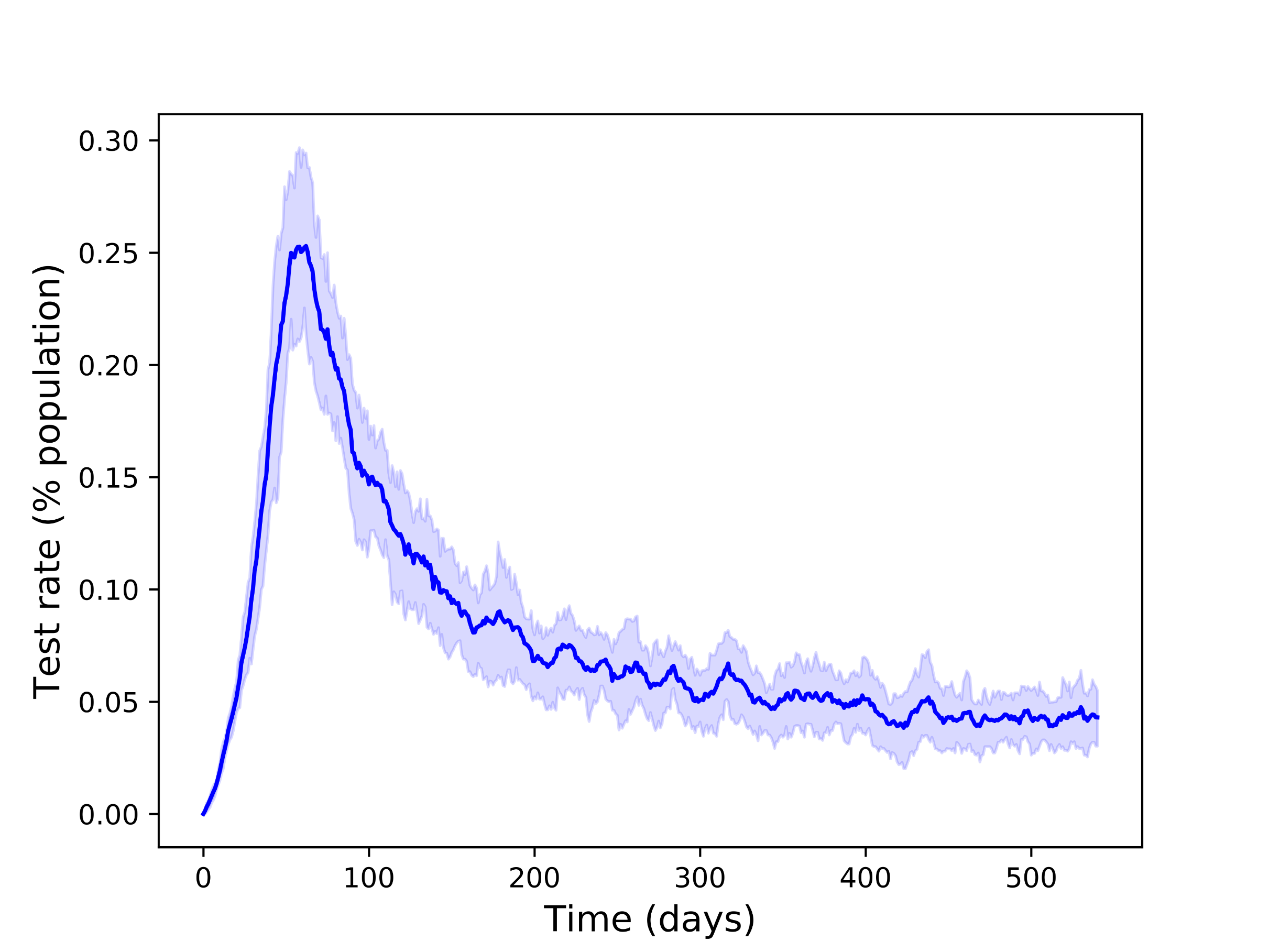}
		\caption{Test rate with time}  
		\label{fig:track_and_test_avg_test_per_day}
	\end{subfigure}
	\begin{subfigure}{.45\linewidth} 
		\includegraphics[width=0.9\linewidth]{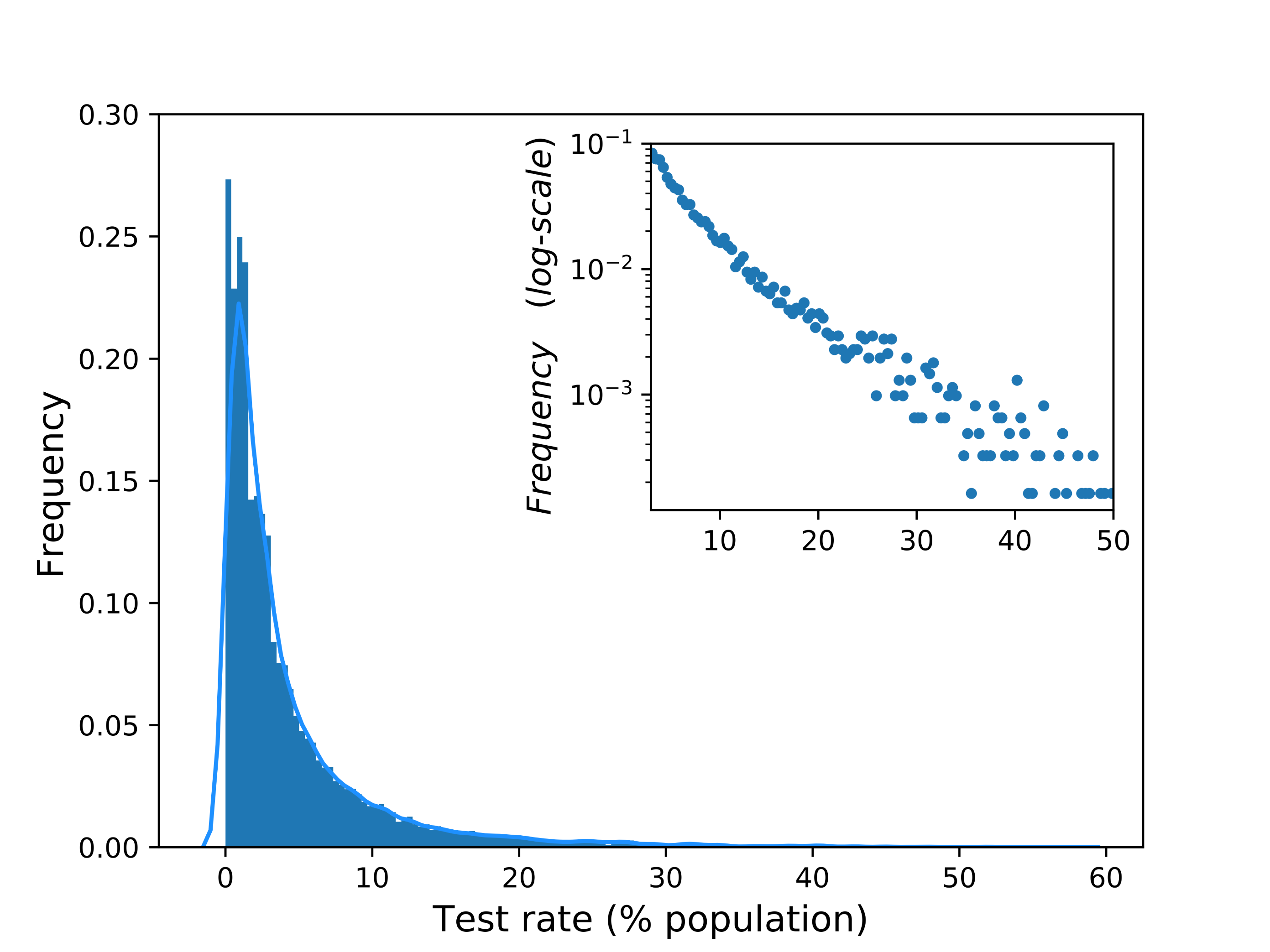}
		\caption{Test rate distribution}  
		\label{fig:track_and_test_test_per_day_hist}
	\end{subfigure}
	\caption{Distribution of test rates of the Track and Test method.}
	\label{fig:track_and_test_test_per_day}
\end{figure}

Figure \ref{fig:track_and_test_avg_test_per_day} shows the average number of tests actually performed by the Track and Test method across time in 50 realizations of the infection spreading process, presented as percentages of the population. The shaded area in the figure shows the 2nd and 3rd quartiles across the different realizations. 
It can be seen that in practice the unbounded Track and Test method typically requires only a moderate number of tests per day.  
The average rate of daily tests increases at the beginning of the outbreak, reaching $0.25\%$ of the population and then gradually decreases.
Figure \ref{fig:track_and_test_test_per_day_hist} shows the full distribution of test rates over all simulation instances and all simulation days. Most of the mass of the distribution is indeed centered at low test rates, but there is a tail of low probability events of days with  high test rates. The inset in logarithmic scale shows that these events are rare and their probability roughly decays exponentially with the test rate, but still, their probability is not negligible. Next we describe a bounded version of the Track and Test method which avoids these high-test-rate events, and show that such test rates are not really required for effective tracking and testing.

\vspace{5pt}
\noindent
\textbf{Bounded Track and Test}: As long as the daily tests did not reach a given capacity limit, the bounded variant of Track and Test runs the recursive tests as in the Track and Test method. If the daily test capacity is reached, bounded Track and Test quarantines all the detected infected individuals and all their neighbors in the tracking network, excluding those that were tested negative on the same day. 
In the special case of zero test capacity, this algorithm is equivalent to the Track and Quarantine algorithm, and in the special case of infinite capacity it is equivalent to the Track and Test algorithm.
\vspace{5pt}

Figure \ref{fig:limited-tests} shows the percentage of total infected population (Figure \ref{fig:limited-tests-infected}) and the economic costs (Figure \ref{fig:limited-tests-cost}) as a function of the daily test capacity in population percentages for the bounded Track and Test method and the Quarantine method. The Quarantine method is combined with random tests with the same test capacity.  
In terms of the size of the outbreak, the bounded version of Track and Test has a significant advantage over the Quarantine method for all test capacities. 
In terms of economic costs, at low test capacities Track and Test has higher costs than the Quarantine method, but the cost drops rapidly as the capacity is increased.  
For test capacities of $3\%$ and above, also the bounded version of Track and Test manages to contain the outbreak without the use of lockdowns and at lower costs than the Quarantine method. 

The above results show that tests and quarantines alone do reduce the size of the infection, but only moderately, and a high testing  capacity is required for a significant reduction in costs and mortality (see also Figure \ref{fig:quarantine-large-scale-tests} in the appendix).
Tracking without tests has a stronger effect, however,  
this is still not sufficient to reach one-digit percentages in the size of the outbreak, and still there are non-negligible costs due to large-scale quarantines. 
The Track and Test method manages to contain the outbreak at low costs and outperforms the other approaches. This advantage of Track and Test is not a result only of a large number of tests, nor a result of tracking alone,  
but it is a result of the combination of tracking with a sufficient testing capacity for contacts of infected individuals.


\subsection{On-Off Lockdown Policies}\label{sec:On-Off}

We have seen that 
on the one hand, in closed systems a strong lockdown 
leads to a fast extinction of the infection, which in turn leads also to low economic costs, but on the other hand, if there are incoming infections the fixed lockdown is not desirable as it leads to high costs. If a sufficient testing capacity is available (of around $3\%$), we have seen that the Track and Test method can effectively contain the outbreak at low costs without lockdowns.
Now we consider a more restrictive setting for interventions, where there is a non-zero incoming flow of infections -- and so a complete eradication of the infection is not possible -- and the daily test capacity is limited to a more moderate value.   
In this setting we turn to evaluate the performance of a family of dynamic lockdown algorithms that we call ``On-Off'' lockdown policies, in the spirit of the ``Hammer and Dance'' idea as advocated in \cite{Pueyo-hammer-and-dance}. 
We evaluate the On-Off lockdown policies in combination with the Track and Test method and with the Quarantine method.

\vspace{5pt}
\noindent
\textbf{On-Off Algorithms:} 
The dynamic ``On-Off'' lockdown policies that we study are a family of threshold algorithms that are defined by four parameters: $\left(Low, High, On, Off\right)$.
The algorithms track the current number of {\em known} active cases, in percentages of the  population size. If this number is above the $On$ threshold, the lockdown level is switched to $High$. If the known active cases are below the $Off$ threshold, the lockdown is switched to $Low$. For example, an On-Off policy with parameters $\left(50, 80, 10, 1\right)$ means that if the number of known active cases is above $10\%$ of the population a strict $80\%$ lockdown will be applied, and once the number of known active cases drops below $1\%$ of the population, the lockdown will be relieved to a $50\%$ level. 
The choice of a lockdown policy is orthogonal to the choice of tracking, testing or quarantine method, in the sense that any On-Off policy can be used in combination with any method of the methods discussed in the previous sections.
\vspace{5pt}

\begin{SCtable}
%
\tiny
\begin{tabular}{lrrrrlrrrr}
\toprule
\multicolumn{5}{c}{Bounded Track and Test} & \multicolumn{5}{c}{Quarantine and random tests} \\
\midrule
 Policy &Low    &  High  &   On   &  Off & \vline \ Policy &Low &High& On & Off\\
\midrule
A   &  0\% & 80\% & 0.02\% & 0.01\% &  \vline \ M &  0\% & 80\% & 0.02\% & 0.01\% \\
B   &  0\% & 80\% & 0.05\% & 0.01\% &  \vline \ N &  0\% & 80\% & 0.05\% & 0.01\% \\
C   &  0\% & 80\% & 0.10\% & 0.01\% &  \vline \ O &  0\% & 80\% & 0.10\% & 0.01\% \\
D   &  0\% & 80\% & 0.20\% & 0.10\% &  \vline \ P &  0\% & 80\% & 0.20\% & 0.10\% \\
E   &  0\% & 80\% & 0.50\% & 0.10\% &  \vline \ Q &  0\% & 80\% & 0.50\% & 0.10\% \\
F   &  0\% & 80\% & 1.00\% & 0.10\% &  \vline \ R &  0\% & 80\% & 1.00\% & 0.10\% \\
\midrule
G   & 50\% & 80\% & 0.02\% & 0.01\% &  \vline \ S & 50\% & 80\% & 0.02\% & 0.01\% \\
H   & 50\% & 80\% & 0.05\% & 0.01\% &  \vline \ T & 50\% & 80\% & 0.05\% & 0.01\% \\
I   & 50\% & 80\% & 0.10\% & 0.01\% &  \vline \ U & 50\% & 80\% & 0.10\% & 0.01\% \\
J   & 50\% & 80\% & 0.20\% & 0.10\% &  \vline \ V & 50\% & 80\% & 0.20\% & 0.10\% \\
K   & 50\% & 80\% & 0.50\% & 0.10\% &  \vline \ W & 50\% & 80\% & 0.50\% & 0.10\% \\
L   & 50\% & 80\% & 1.00\% & 0.10\% &  \vline \ X & 50\% & 80\% & 1.00\% & 0.10\% \\
\bottomrule
\end{tabular}
\caption{On-Off policies corresponding to the results in Figure \ref{fig:On-Off}. The left columns are On-Off policies combined with the Track and Test method with a bounded daily test capacity of $0.5\%$. The right columns are with the Quarantine method and the same number of tests distributed at random. The middle line separates between the``strict'' and ``trembling-hand'' On-Off policies.}\label{fig:table-On-Off}
\end{SCtable}

Figure \ref{fig:on-off-map} shows the cost map of 24 policies that are a combination of 12 On-Off algorithms applied with two methods: the bounded version of Track and Test and the Quarantine method with random tests, both with a daily test capacity of $0.5\%$ of the population. The letters in the figure mark the different policies as shown in Table \ref{fig:table-On-Off}; the 12 left columns in the table are with the bounded Track and Test method and the right columns are with the Quarantine method. 
Larger circles in Figure \ref{fig:on-off-map} indicate higher variance between simulation instances and the radius shows the standard deviation.
The purple and orange dotted curves are 
the results of different fixed lockdown policies with the same number of tests for Track and Test (purple) and for Quarantine (orange), presented for comparison with the On-Off policies.

The On-Off policies are divided to two types: policies that switch between a $80\%$ lockdown and $0\%$ lockdown, which we call ``strict'' On-Off policies, and policies that switch between $80\%$ and $50\%$ lockdowns, which we call ``trembling-hand'' On-Off policies.
For the Quarantine method, policies $M$-$R$ are strict and policies $S$-$X$ are trembling-hand. Each of these two families shows a cost-mortality trade-off: changes in the thresholds of the algorithm within the same family that reduce mortality also increase the economic cost. When switching between the trembling-hand family $S$-$X$ to the strict family $M$-$R$ the trade-off breaks and it is possible by such changes in the algorithm to reduce both costs and mortality, except for policy $S$ that has a slightly lower mortality than policy $M$.
For the Track and Test method (purple markers in the figure), policies $A$-$E$ are strict and policies $G$-$L$ are trembling-hand On-Off policies. Policies $G$-$L$ all have similar high cost and low mortality, and policies $A$-$E$ show the cost-mortality trade-off as in the Quarantine method. 

An especially interesting policy in Figure \ref{fig:on-off-map} is policy $A$ that achieves a very low mortality rate and has a moderate cost of $35\%$, significantly lower than the fixed lockdown alternatives (the dotted lines in the figure). Figure \ref{fig:on-off-A-dynamics} shows the dynamics of the spread of the infection under policy $A$. It can be seen that under this policy the $80\%$ lockdown is active less than half of the time, and is restarted at a frequency of about one month. This algorithm leads to a slow quasi-linear spread of the infection, reaching less than $2\%$ of the population in $18$ months.

We conclude that first, the dynamic On-Off policies are more efficient than fixed lockdown policies, and second, also in the setting of a moderate testing capacity and in combination with dynamic lockdowns, Track and Test leads to a substantial reduction in mortality and costs. 
Regarding the choice between specific types of On-Off policies, first, we see that trembling-hand policies have only a minor advantage in mortality rate over strict On-Off policies, but cause a significantly higher economic burden, and second, the low threshold (and hence high frequency) policies achieve better results in terms of mortality than the high-threshold ones, but with some trade-off between mortality and economic cost.

\section{Conclusion}
We have seen examples of the basic trade-off between economic and mortality costs, where different intervention methods present different versions of this trade-off. The choice of a policy involves multiple components such as the number of tests, the method in which tests are distributed, the level of social distancing restrictions and whether this level is static or dynamic, whether tracking is applied or not, and the method in which the tracking information is used. Every policy eventually tries to trade-off between economic costs and mortality, as well as privacy costs and other social costs incurred by the epidemic or by the policy itself.
 
In presenting policy outcomes as points on the economic-mortality cost map we have shown that there are clear differences between policy types, which transcend the basic trade-off in the sense of Pareto domination. Specifically, our results show that Track and Test policies dominate non-tracking policies, reaching low mortality and economic costs. 
If the testing capacity is around  $3\%$, Track and Test manages to stop the spread of an infection, without lockdowns. 
When the testing capacity is lower, we have seen that On-Off lockdowns dominate fixed-lockdown policies, and again, Track and Test policies dominate non-tracking policies. 

Our findings on the benefits of fast and effective tracking of all infected individuals and their interactions suggest that these have the potential to replace the usage of lockdowns and of social distancing measures, and to save human lives and economic costs.
Digital tracking has a serious cost in the privacy of infected individuals and their contacts, but on the other hand, lockdowns and social distancing restrictions have their own costs in freedom and human rights, beyond their economic costs. 
Our results stress the importance and urgency of facing the difficult normative question of whether and how to deploy contact tracing in democratic countries, both for the current COVID-19 crisis and in order to prevent outbreaks in the future.

\bibliographystyle{splncs03}
\bibliography{contact_tracing_and_testing_SEIR}

\newpage
\appendix
\section{Geographic Small World Networks} \label{sec:geo-graphs}
In the main text we present results with dynamic random graphs that are similar to the classical SEIR model, assuming uniform mixing among the population. Here we present results of experiments with a network model that takes into account geographic distances as well as small-world network effects \cite{kleinberg2000small}. The model assumes that vertices are located on a two-dimensional grid and interact with all their geographically close neighbors. In addition, there is a probability for each vertex to have a long-distance interaction, where the probability is inversely proportional to the square of the distance.  For more details of the geographic small world network model see \cite{kleinberg2000small}. 
Here we show that the results with these networks are very similar to the results presented in the main text, indicating that even when the population mixing is not uniform but is strongly restricted by geographic distances, the basic effects of different intervention approaches and the relations between these approaches remain the same.

\subsection{Fixed Lockdowns}
\vspace{-10pt}
\begin{figure}[h!]
\centering
	\begin{subfigure}{.45\linewidth}
		\includegraphics[width=0.9\linewidth]{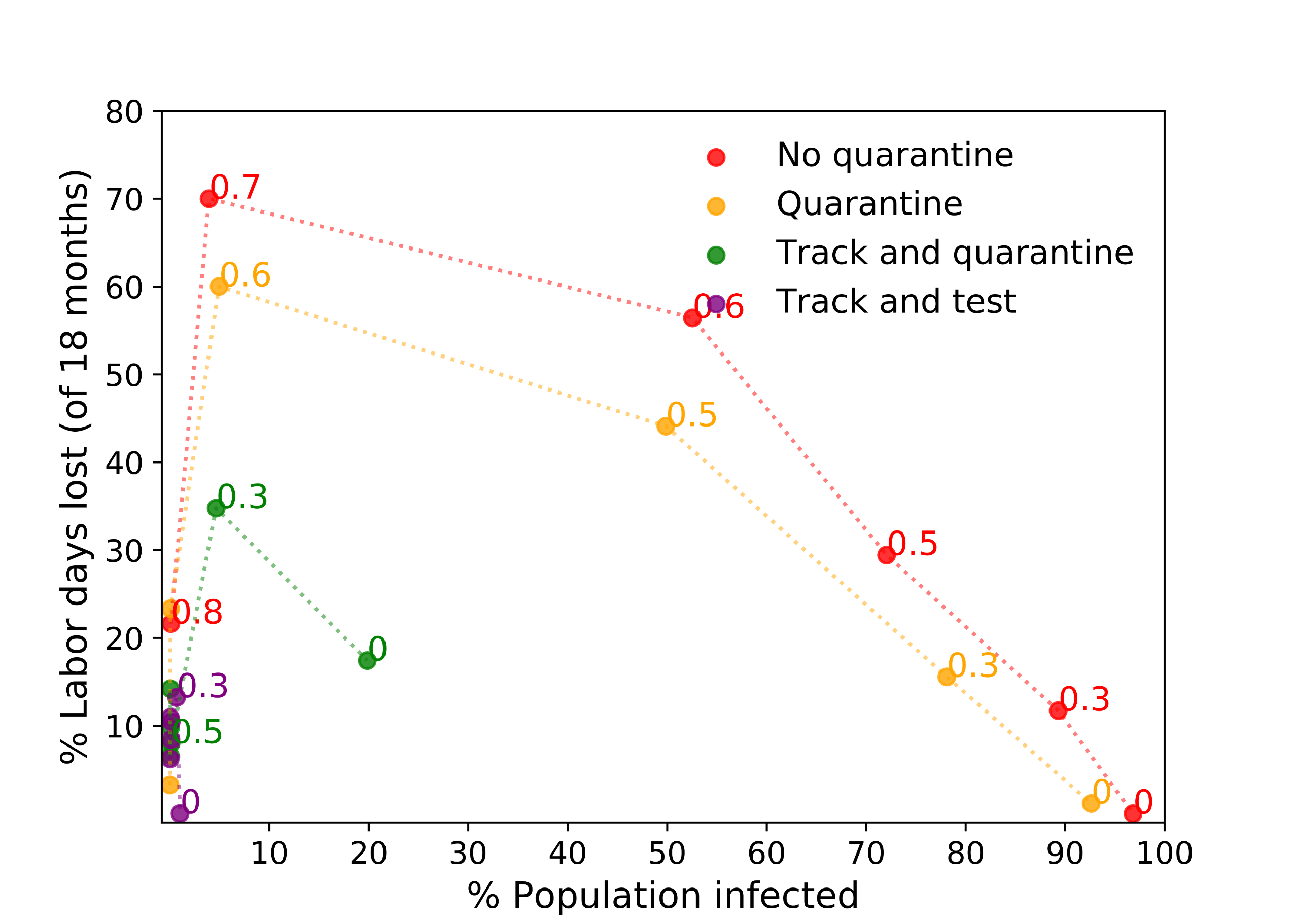}
		\caption{Closed system}  
		\label{fig:geo_graphs_fixed_lockdown_no_flow}
	\end{subfigure}
	\begin{subfigure}{.45\linewidth} 
		\includegraphics[width=0.9\linewidth]{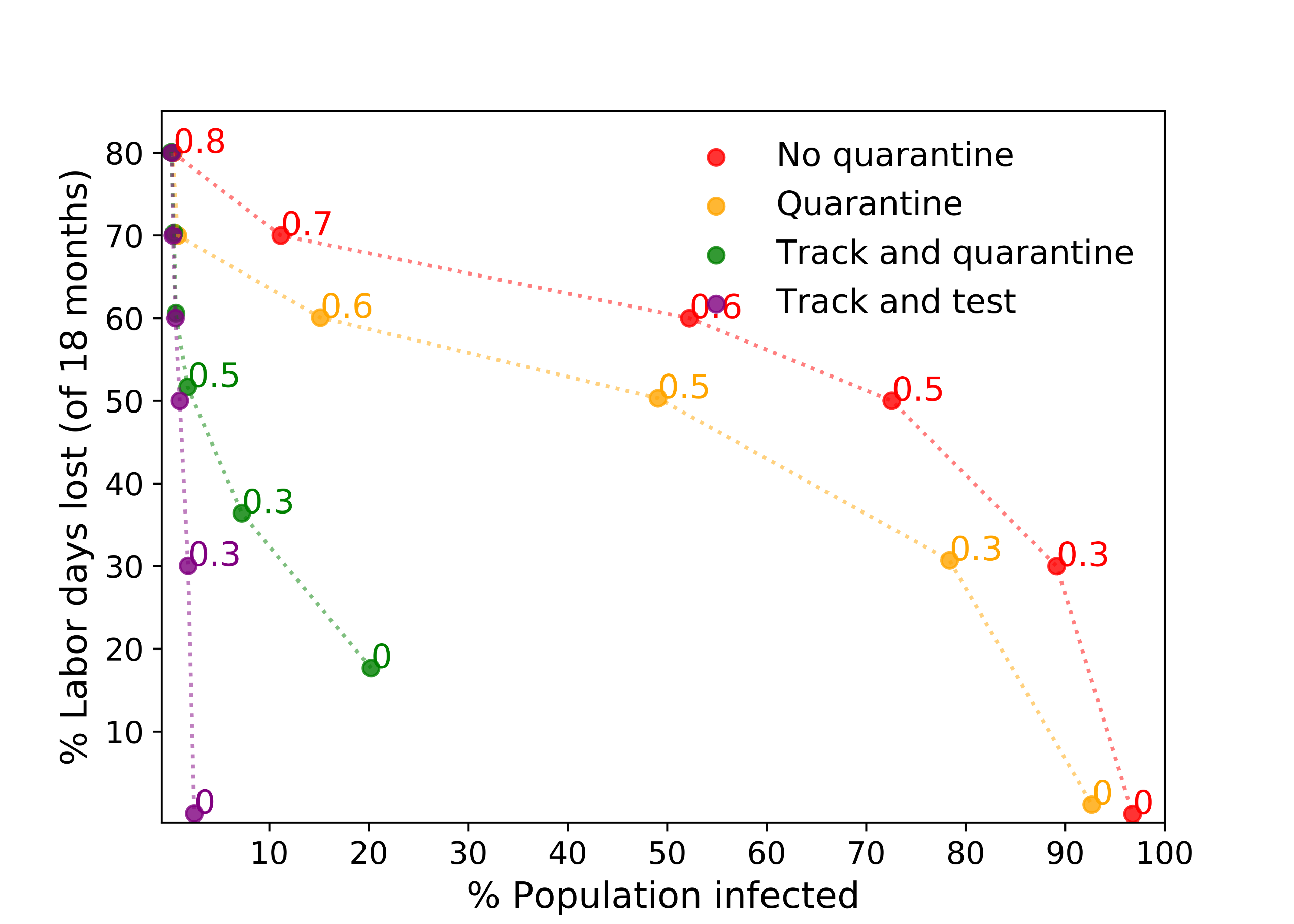}
		\caption{Open system}  
		\label{fig:geo_graphs_fixed_lockdown_with_flow}
	\end{subfigure}
	\caption{Tracking compared with the Quarantine method and the no-quarantine benchmark with different levels of fixed lockdowns in geographic small world networks.}
	\label{fig:geo_graphs_fixed_lockdown}
\end{figure}

\subsection{On-Off Lockdown Policies}
\vspace{-10pt}
\begin{figure}[h!]
\centering
	\begin{subfigure}{.45\linewidth}
		\includegraphics[width=0.9\linewidth]{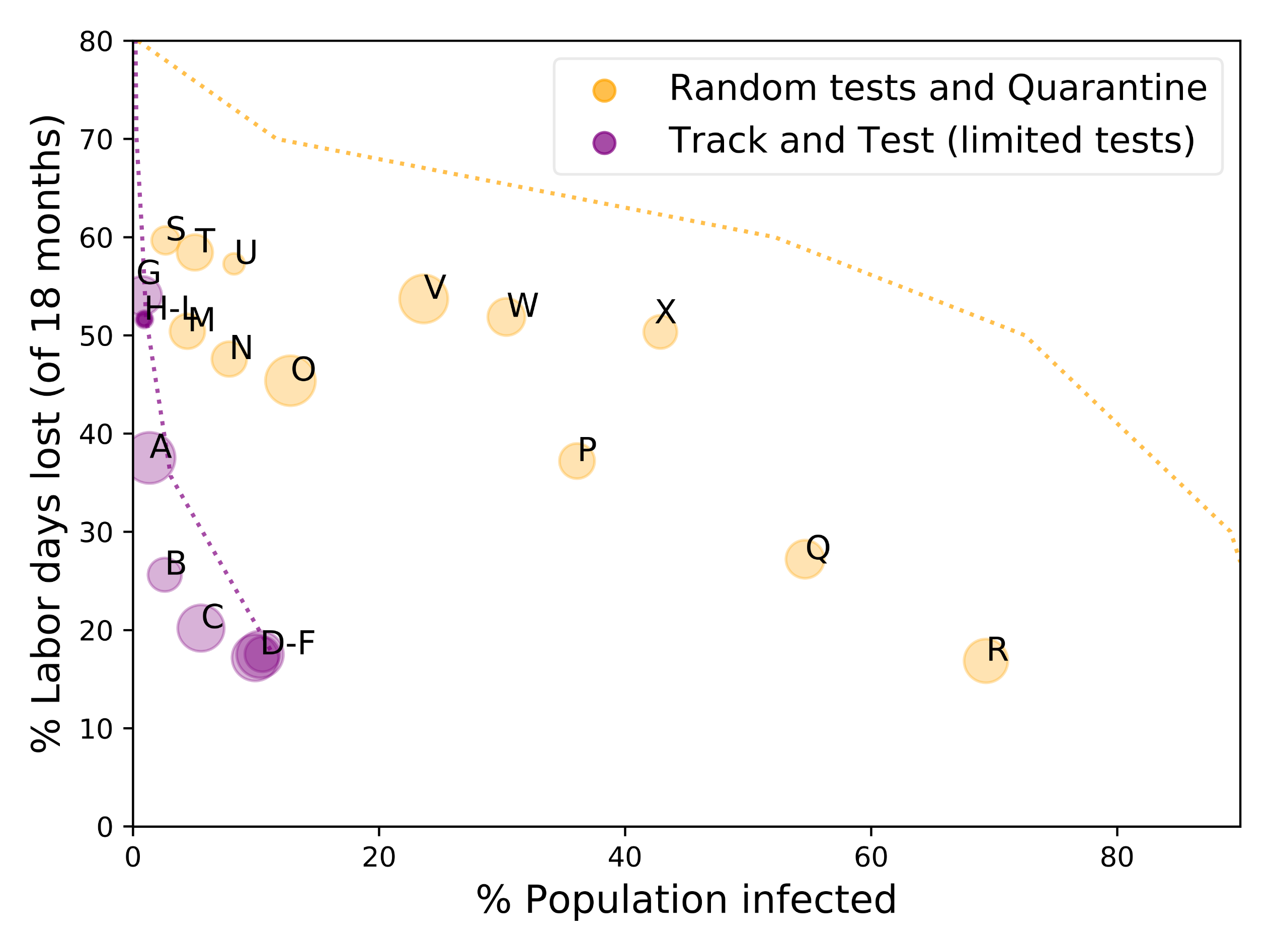}
		\caption{}  
		\label{fig:on_off_geo_graphs}
	\end{subfigure}
	\begin{subfigure}{.45\linewidth}
	\vspace{-10pt}
		\includegraphics[width=0.935\linewidth]{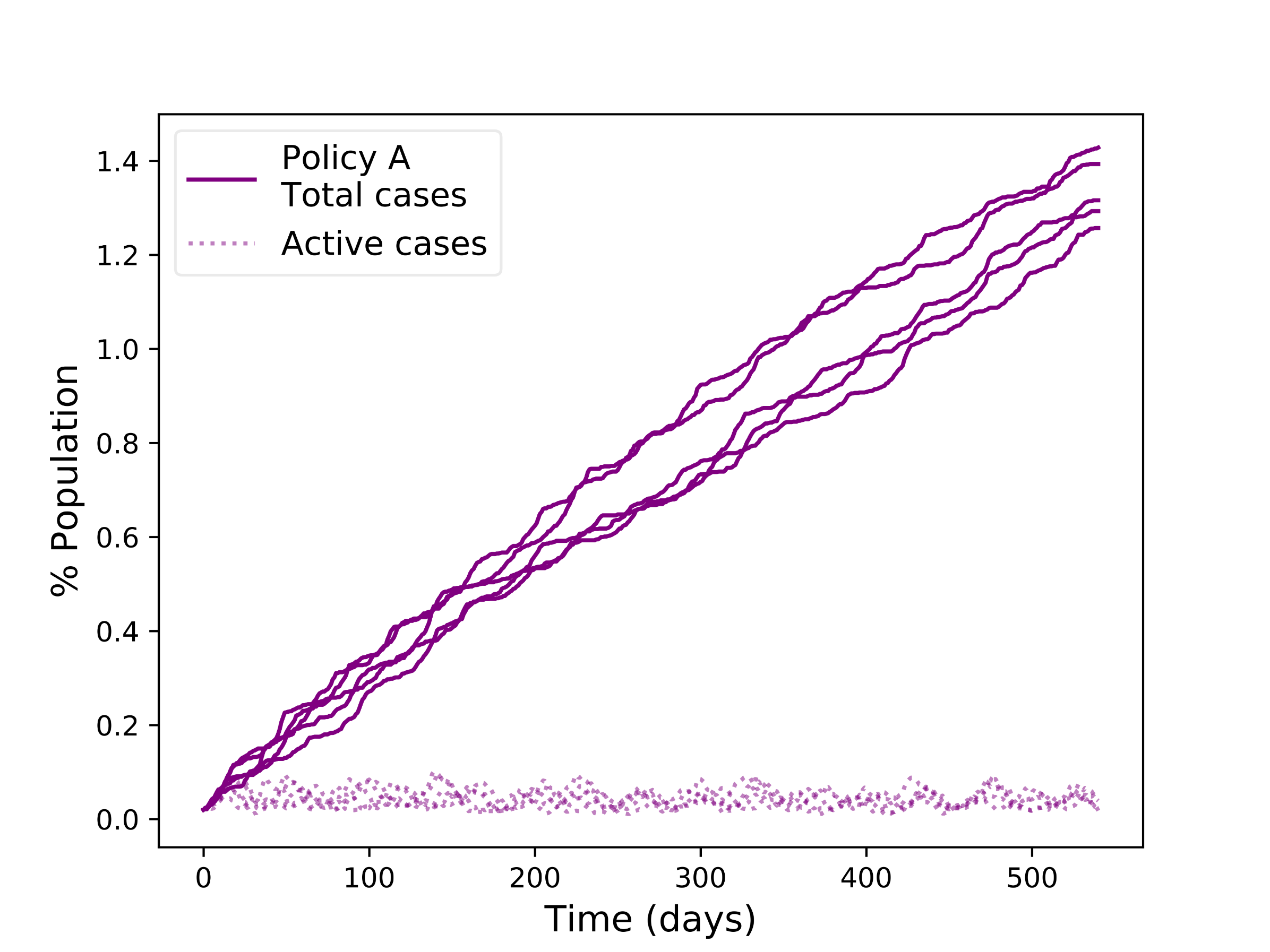}
		\caption{}  
		\label{fig:on-off-A-dynamics-geo-graphs}
		\vspace{4pt}
	\end{subfigure}
\caption{On-Off policies in geographic small world networks. \ref{fig:on-off-map}: The cost map of the total percentage of labor days lost in $18$ months versus the total percentage of population infected. The letters near the markers indicate the different policies as shown in Table \ref{fig:table-On-Off}. Purple markers are policies with the bounded Track and Test method and orange markers are with the Quarantine method with random tests. The purple and orange dotted curves show a comparison to the results of fixed lockdown policies with bounded Track and Test and with Quarantine and random tests, respectively. \ref{fig:on-off-A-dynamics}: Dynamics of the outbreak under policy $A$ -- an example of five simulation instances.}
\label{fig:geo-graphs-On-Off}
\end{figure}

\section{Quarantine with Large Scale Testing}
\vspace{-10pt}
\begin{figure}[h!]
\centering
	\begin{subfigure}{.45\linewidth}
		\includegraphics[width=0.9\linewidth]{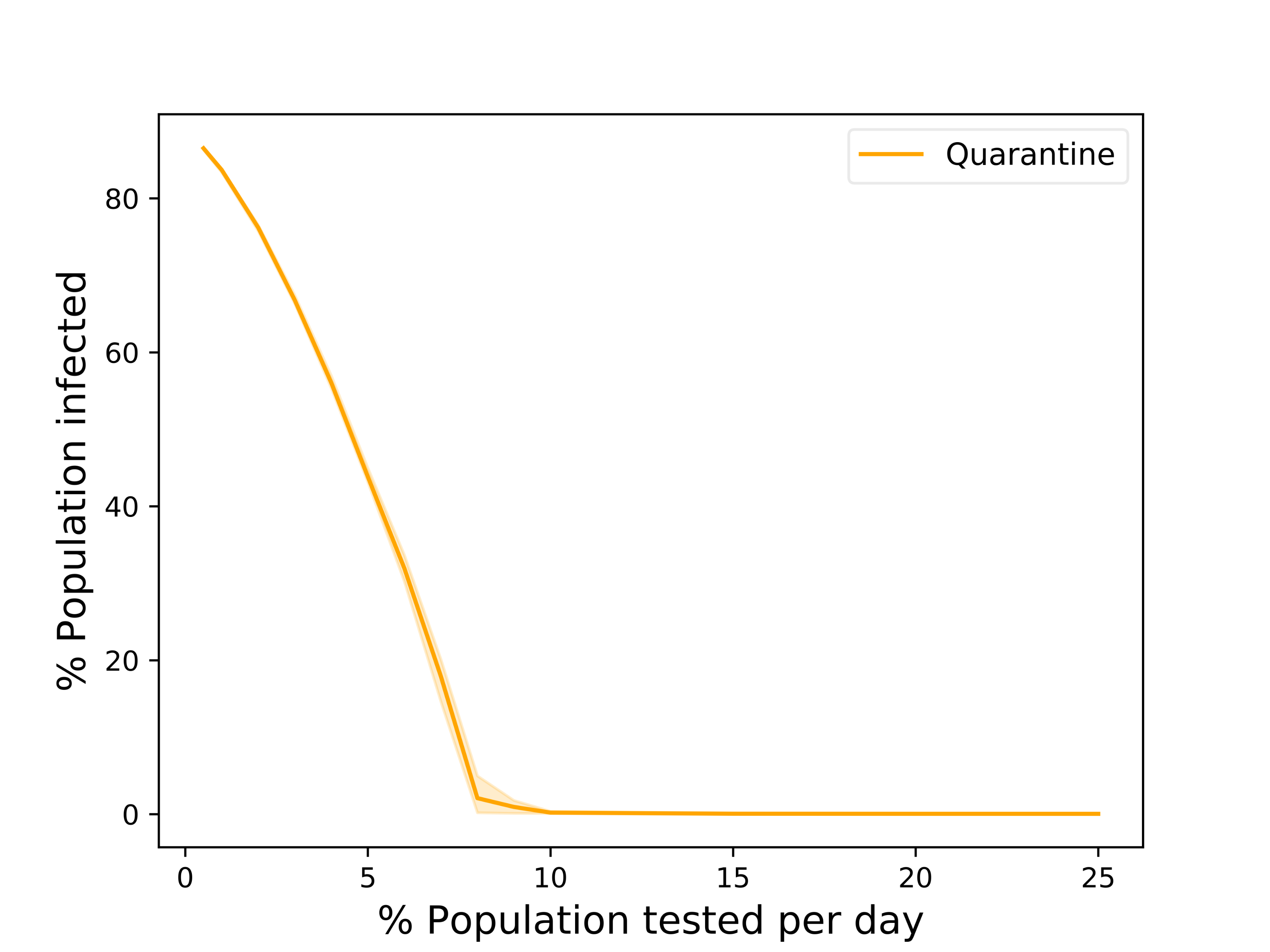}
		\caption{Total infected}  
		\label{fig:quarantine-large-scale-tests-mortality}
	\end{subfigure}
	\begin{subfigure}{.45\linewidth} 
		\includegraphics[width=0.9\linewidth]{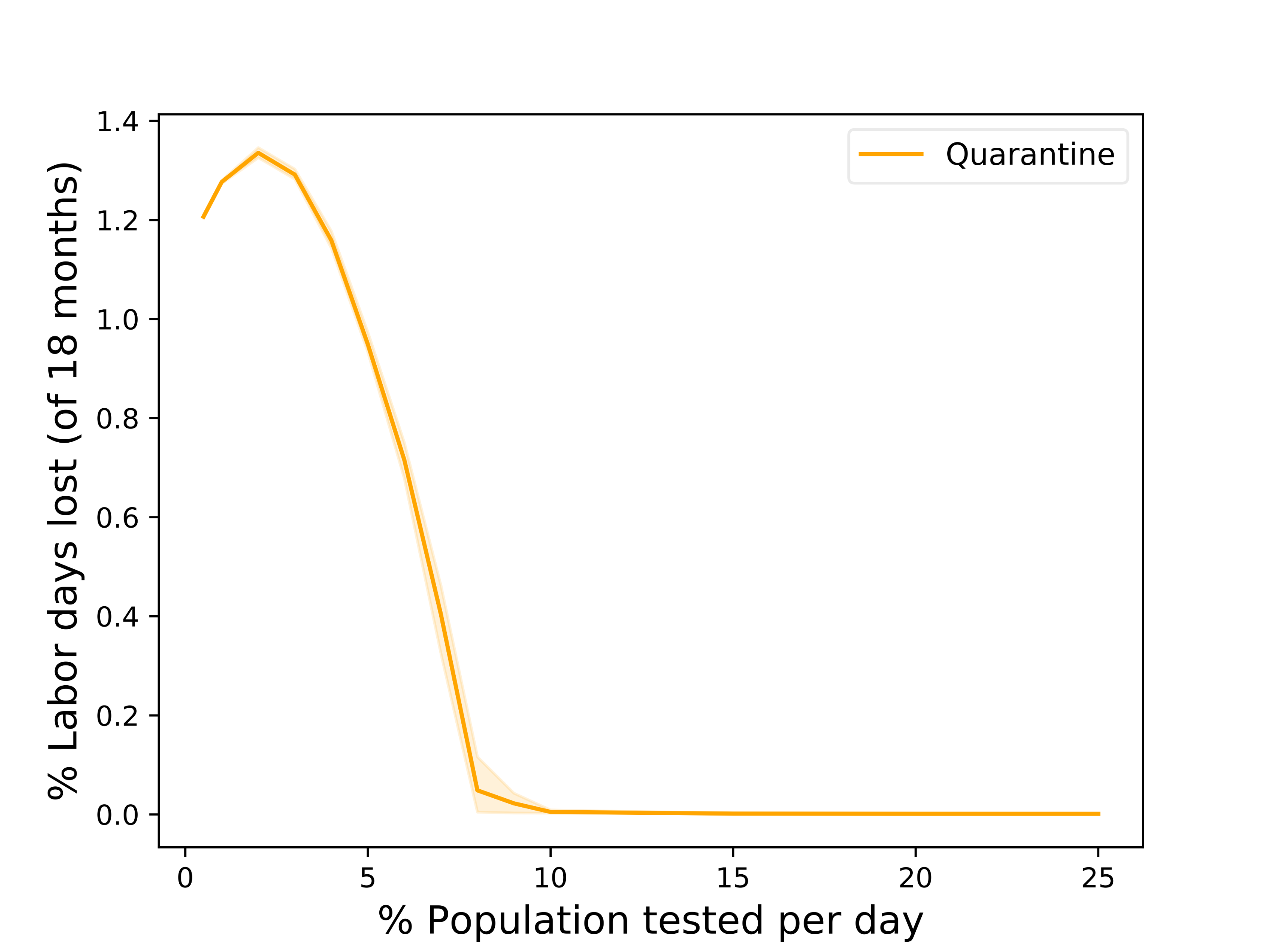}
		\caption{Economic costs}  
		\label{fig:quarantine-large-scale-tests-cost}
	\end{subfigure}
	\caption{Effect of large-scale testing: The Quarantine method combined with random tests with high testing capacities. Daily testing capacities are presented in percentages of the population. 
	\ref{fig:quarantine-large-scale-tests-mortality}: Total percentage of population infected as a function of the testing capacity.
	\ref{fig:quarantine-large-scale-tests-cost}: Economic costs as a function of the testing capacity.   
	}
	\label{fig:quarantine-large-scale-tests}
\end{figure}


\end{document}